\documentclass[]{aa}
\usepackage{graphicx}
\usepackage{amssymb}
\usepackage[normalem]{ulem}
\newcommand{\pheins}{\phantom{1}}   
\newcommand{\wav}[1]{$\lambda#1\,{\rm cm}$} 
\newcommand{\wwav}[2]{$\lambda\lambda#1,#2\,{\rm cm}$}  
\newcommand{\wwwav}[3]{$\lambda\lambda#1,#2,#3\,{\rm cm}$}  
\newcommand{\Bt}{\,B_\mathrm{tot}}  
\newcommand{\Btpe}{\,B_\mathrm{tot\perp}}   
\newcommand{\Br}{\,B}  
\newcommand{\Brpa}{\,B_{\parallel}}  
\newcommand{\Brpe}{\,B_{\perp}}  
\newcommand{\Bur}{\,B_{r}}  
\newcommand{\But}{\,B_{\theta}}   
\newcommand{\Buz}{\,B_{z}}  
\newcommand{\bt}{\,b}  
\newcommand{\btpe}{\,b_{\perp}}  
\newcommand{\cm}{\,{\rm cm}}

\newcommand{\cmcube}{\,{\rm cm^{-3}}}

\newcommand{\FRM}{\,{\rm rad\,m^{-2}}}

\newcommand{\kpc}{\,{\rm kpc}}

\newcommand{\mkG}{\,\mu{\rm G}}

\newcommand{\p}{\,{\rm pc}}
\newcommand{\radm}{\,{\rm rad\,m^{-2}}}

\newcommand{\RM}{\,\mathrm{RM}}
\newcommand{\RMi}{\,\mathrm{RM_i}}
\newcommand{\RMfg}{\,\mathrm{RM_{fg}}}

\newcommand{\disk}{_\mathrm{d}}
\newcommand{\DP}{\mathrm{DP}}

\newcommand{\screen}{_\mathrm{s}}
\newcommand{\syn}{_\mathrm{syn}}
\newcommand{\therm}{_\mathrm{th}}

\begin{document}


\title{The magnetic field of \object{M\,31} from multi-wavelength
  radio polarization observations}

\author{A.\,Fletcher\inst{1}
  \and E.M.\,Berkhuijsen\inst{1}
  \and R.\,Beck\inst{1}
  \and A.\,Shukurov\inst{2}}

\offprints{A.\,Fletcher, \\
  \email{fletcher@mpifr-bonn.mpg.de}}

\institute{Max-Planck-Institut f{\"u}r Radioastronomie, Auf dem
  H{\"u}gel 69, D-53121 Bonn, Germany \and School of Mathematics and Statistics,
  University of Newcastle, Newcastle upon Tyne, NE1 7RU, UK}
\date{Received / Accepted}
\abstract{The configuration of the regular magnetic field in M\,31 is
  deduced from radio polarization observations at the wavelengths
  $\lambda\lambda\ 6,\ 11$ and $20\cm$. By fitting the observed
  azimuthal distribution of polarization angles, we find that the
  regular magnetic field, averaged over scales 1--3\,kpc, is almost
  perfectly axisymmetric in the radial range $8$ to $14\kpc$, and
  follows a spiral pattern with pitch angles of $p\simeq -19\degr$ to
  $p\simeq -8\degr$. In the ring between $6$ and $8\kpc$ a
  perturbation of the dominant axisymmetric mode may be present,
  having the azimuthal wave number $m=2$.  A systematic analysis of
  the observed depolarization allows us to identify the main mechanism
  for wavelength dependent depolarization -- Faraday rotation measure
  gradients arising in a magneto-ionic screen above the synchrotron
  disk. Modelling of the depolarization leads to constraints on the
  relative scale heights of the thermal and synchrotron emitting
  layers in M\,31; the thermal layer is found to be up to three times
  thicker than the synchrotron disk. The regular magnetic field must
  be coherent over a vertical scale at least similar to the scale
  height of the thermal layer, estimated to be $h\therm\simeq 1\kpc$.
  Faraday effects offer a powerful method to detect thick
  magneto-ionic disks or halosaround spiral galaxies.
 }
\titlerunning{The magnetic field in M\,31}
\authorrunning{A.\,Fletcher et al.}

\maketitle

\section{Introduction}
\label{sec:intro}

The Andromeda nebula, M\,31, is the nearest spiral galaxy to the Milky
Way. Despite its high inclination to the line of sight, the large
angular size of the galaxy allows detailed studies of its magnetic
field and interstellar medium (ISM). In particular, the large scale
morphology of the magnetic field can be investigated with unmatched
precision. M\,31 is thus of prime importance in bringing together
observational data and theory about galactic magnetic fields.

Early radio wavelength observations of M\,31 at \wav{73} (Pooley
\cite{Pooley69}) and \wav{11} (Berkhuijsen \& Wielebinski
\cite{Berkhuijsen74}, Berkhuijsen \cite{Berkhuijsen77}) show the
continuum emission concentrated in a ring, at a radius of $r\simeq
50\arcmin \simeq 10\kpc$. The first radio polarization observations at
\wav{11}, using the 100m Effelsberg telescope (Beck et al.\ 
\cite{Beck78}), indicated that the magnetic field in the southern part
of M\,31 is aligned with the optical spiral arms. Beck (\cite{Beck82})
interpreted the \wav{11} data by comparing the observed polarization
angles with a model of the polarized emission to reveal a
predominantly azimuthal large-scale magnetic field, concentrated in
the $r\simeq 10\kpc$ `ring', directed in the same direction as the
rotation of the galaxy.  Faraday rotation measures (RMs) from
polarization observations of the southwestern arm of M\,31 at
\wwav{6}{20} confirmed the presence of a basically axisymmetric spiral
magnetic field (Beck et al.\ \cite{Beck89}). A bisymmetric component
of the magnetic field was suggested by Sofue \& Beck (\cite{Sofue87})
from an analysis of the deviation of the polarization angles at
\wav{11} from those expected due to a purely axisymmetric regular
magnetic field; however it is not clear whether the inferred
bisymmetric mode is statistically significant. Ruzmaikin et al.\ 
(\cite{Ruzmaikin90}) modelled the \wav{11} polarization angles of
M\,31 with an azimuthal Fourier expansion for the regular magnetic
field and ascertained that deviations of the magnetic field from axial
symmetry are evident statistically and may indicate bisymmetric or
higher modes. More recently, RMs of 21 background radio sources in the
field of M\,31 were found to be compatible with the same magnetic
field structure, but extending far away from the $r\simeq 10\kpc$
`ring', probably to $5\lesssim r \lesssim 25 \kpc$ (Han et al.\ 
\cite{Han98}).  This remains to be substantiated with a statistically
significant number of sources.

Recently, Berkhuijsen et al.\ (\cite{Berkhuijsen03}) presented a new
\wav{6} survey of M\,31 and concluded that: the regular component of
the magnetic field is probably as strong as the turbulent field; the
regular magnetic field has an average pitch angle of $\simeq -15\degr$
in the range $8\lesssim r \lesssim 12\kpc$, with a negative value
indicating a trailing spiral; gradients in Faraday rotation measure
may be an important cause of depolarization.

In this paper we seek to take the next logical step in understanding
the magnetic structure of M\,31 by developing a detailed and
self-consistent description of the magnetic field. We use \emph{all}
of the radio polarization surveys (\wwwav{6}{11}{20}) and fit together
information on polarization angles, Faraday rotation, non-thermal
radio emission intensities, depolarization and the scale heights of
ISM components. Our analysis has two main components: deducing the
large-scale geometry of the magnetic field and deriving parameters of
the magneto-ionic ISM from analysis of depolarization of the
synchrotron emission. Our approach is the latest in a sequence of
methods used to interpret radio polarization observations of external
galaxies. Ruzmaikin et al.  (\cite{Ruzmaikin90}) considered the
variation of polarization angles at a single wavelength, Sokoloff et
al. (\cite{Sokoloff92}) extended this approach to multiple wavelengths
and Berkhuijsen et al.  (\cite{Berkhuijsen97}) introduced variation in
the intrinsic angle of polarized emission in a galaxy. We develop a
new model, by combining an analysis of multi-wavelength polarization
angles -- based on the earlier methods -- with modelling of the
wavelength dependent depolarization.

A short description of the data we use is presented in
Sect.~\ref{sec:data}.  The properties of the synchrotron disk are
discussed in Sect.~\ref{sec:nthdisk}. In Sect.~\ref{sec:model} we use
polarization angles at \wwwav{6}{11}{20} to deduce the
three-dimensional structure of the regular magnetic field in M\,31.
The method, developed from that used by Berkhuijsen et al.\
(\cite{Berkhuijsen97}) to determine the regular magnetic field of
M\,51, takes into account the intrinsic angle of polarized emission in
the disk of M\,31, Faraday rotation by the magneto-ionic medium in
M\,31 and Faraday rotation in the Milky Way. In
Sect.~\ref{sec:depolar} we analyze the radial and azimuthal variation
in the depolarization between wavelengths \wav{6} and \wav{20}, and
derive constraints on the scale heights of the thermal and synchrotron
emitting disks of M\,31.  This demonstrates a new and potentially
powerful method for extracting such information from radio
polarization observations of spiral galaxies. A short discussion of the
preliminary results was presented in Fletcher et al.\ (\cite{Fletcher00}).

\section{Observational data}
\label{sec:data}

\subsection{Radio continuum emission at $\lambda\lambda$6, 11 and 20\,cm}
\label{subsec:maps}
\begin{figure*}
  \centerline{
    \includegraphics[width=0.95\textwidth,angle=0]{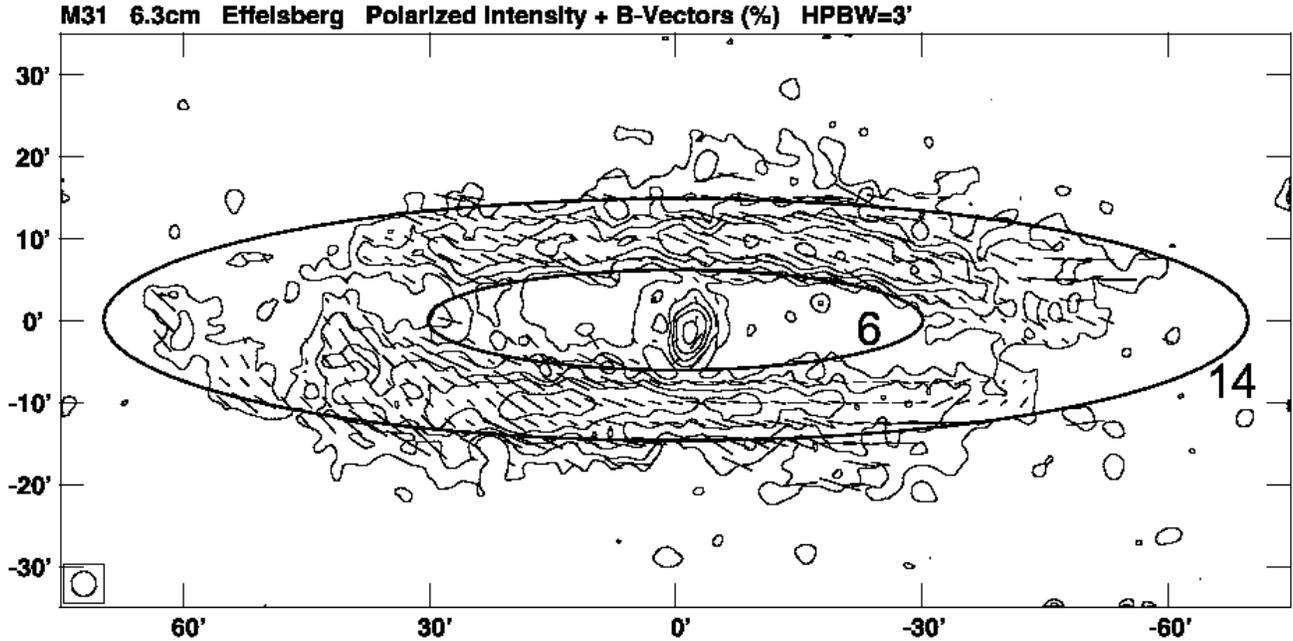}}
  \caption{Polarized intensity (contours) of M\,31 with the
    orientation of the emission B-vector also shown (dashes, not
    corrected for Faraday rotation) with their lengths proportional to
    the degree of polarization, observed at \wav{6} with the
    Effelsberg radio telescope (Berkhuijsen et
    al.~\cite{Berkhuijsen03}). Note that the foreground RM of $-90\FRM$
  (Table 2) corresponds to Faraday rotation of about $20\degr$ so
that the intrinsic B-vectors are roughly azimuthal.
    The beam width is $3\arcmin$ and the rms noise is 0.2~mJy/beam.
    Contour levels are $1,2,3,4,6 \times(5\times 10^{-4})$\,Jy/beam. A
    length of B-vectors of $3\arcmin$ corresponds to a degree of
    polarization of 36\%. The northern major axis is to the left and
    the ellipses show the radial range of the data analyzed in this
    paper, $6\le r\le14\kpc$.}
  \label{fig:PI}
\end{figure*}

For the analysis in this paper we adopt the following parameters of
M\,31: a distance of $690\kpc$ ($1\arcmin = 196\p$ on the major axis),
a centre position of $\rm RA_{50} = 0^h 40^m 1\fs 8,\ DEC_{50} =
40\degr 59\arcmin 46\arcsec$, an inclination angle of $i=78\degr$
(Braun\ \cite{Braun91}) where $0\degr$ is face-on, and a position
angle of the northern major axis of $37\degr$.

Berkhuijsen et al.\ (\cite{Berkhuijsen03}) observed a field of
$150\arcmin \times 70\arcmin$ at \wav{6.2} with the 100m Effelsberg
telescope . The original resolution was $2\farcm4$.
Figure~\ref{fig:PI} shows the polarized intensity smoothed to a
beamwidth of $3\arcmin$, along with ellipses showing the radial range
considered in this paper.  Preliminary results were also discussed by
Han et al.\ (\cite{Han98}) and Beck (\cite{Beck00}).

The \wav{11.1} map of M\,31 was obtained with the Effelsberg telescope
and was published by Beck (\cite{Beck82}). The original resolution of
$4\farcm4$ was smoothed to $5\arcmin$ for this paper. The VLA map at
\wav{20.5} by Beck et al.\ (\cite{Beck98}) has an original resolution of
$45\arcsec$, used in the analysis of polarization angles in
Sect.~\ref{sec:model}. For the comparison of polarized intensities
at \wwav{6}{20} presented in Sect.~\ref{sec:depolar}, the \wav{20}
map was smoothed to $3\arcmin$, the same as the \wav{6} map. Since we
are interested in \emph{wavelength dependent} effects (Faraday
depolarization) we require the same degree of \emph{wavelength
  independent} depolarization at \wav{6} and \wav{20} and so the
resolutions must be the same (wavelength independent depolarization
arises from unresolved fluctuations of the polarized emission -- see
Appendix~\ref{app}).

From the \wav{20} map in total intensity, smoothed to a resolution of
$3\arcmin$, and the \wav{6} total intensity map at the same resolution
Berkhuijsen et al.\ (\cite{Berkhuijsen03}) computed a spectral index
map and maps of thermal and non-thermal emission at these wavelengths.
Combination of the polarized and non-thermal emission at each
wavelength then yields the non-thermal degrees of polarization that
are analyzed in Sect.~\ref{sec:depolar}.

Missing spacings affect the diffuse emission at \wav{20} detectable by
the VLA. This was corrected in total intensity with the help of
Effelsberg data at the same wavelength. A further complication is that
M\,31 lies behind a spur of the Milky Way seen in \wav{20} non-thermal
emission (see., e.g., Gr\"ave et al.\ \cite{Graeve81}, Beck et al.\ 
\cite{Beck98}). The total emission from the foreground at \wav{20} was
removed using an Effelsberg map of the extended region of sky in the
direction of M\,31, but the polarized emission cannot be separated
into Milky Way and M\,31 components so that a correction for missing
spacings was not possible in the maps of Stokes $Q$ and $U$ (Beck et
al.\ \cite{Beck98}).  However, strong spatial variation in the Faraday
rotation intrinsic to M\,31, shown in Figure~\ref{fig:RM} for RM
derived using \wwwav{6}{11}{20} data means that at \wav{20} Stokes $Q$
and $U$ originating from M\,31 will change rapidly with position and
hence the effect of missing spacings is probably small for the
emission from M\,31. Note that a similar pattern is present when RM is
determined using only \wwav{6}{11} (Fig.~12 in Berkhuijsen et
al.~\cite{Berkhuijsen03}).

\subsection{Data averaging in rings and sectors}
\label{subsec:averaging}

The maps in the Stokes parameters $I$, $Q$ and $U$, at each of the three
wavelengths, were averaged in sectors of $20\degr$ azimuthal and $2\kpc$
radial width, in the range $6\le r\le 14$. The size of the sectors was chosen
to match the resolution of the data at \wav{6}. Next we describe how the
average $Q$ and $U$ intensities in each sector were combined to give the
average polarization angle and the average polarized emission intensity in
each sector.

\subsubsection{Polarization angles}
\label{subsubsec:pa}

The polarization angle in a individual sector was calculated as
$\psi=\frac{1}{2}\arctan{\langle U\rangle}/{\langle Q\rangle}$, where
$\langle \dots\rangle$ denotes the average value of the parameter over
the pixels within a sector. The resolutions used were $3\arcmin$,
$5\arcmin$ and $45\arcsec$ at the wavelengths \wwwav{6}{11}{20}
respectively. The errors in polarization angle were computed as the
standard deviations, within one sector, between all pixels whose
intensity is stronger than three times the rms noise level. If the
number of pixels in a sector was below five, the error was calculated
by averaging several adjacent sectors (this procedure was suggested by
Berkhuijsen et al.\ \cite{Berkhuijsen97}). For two measurements (both
at \wav{20}, in the ring 6--8 kpc at $\theta=120\degr$ and in the ring
8--10 kpc at $\theta=60\degr$) the error thus obtained was less than
the noise in the maps and here the noise error was taken.  These
average polarization angles are analysed in Sect.~\ref{sec:model}.

The \wav{20} polarized emission from M\,31 is mixed with a substantial
amount of emission from the Milky Way foreground. At $45\arcsec$
resolution the polarized emission from the M\,31 `ring' and nucleus is
clearly visible and the polarization angles are clustered in coherent
cells, sometimes connected with the position of OB associations in
M\,31 (see Figs.~2 and 6 of Beck et al.\ \cite{Beck98}). Thus, the
average \wav{20} polarization angles in sectors with a surface area
several tens of times larger than the $45\arcsec$ resolution, are a
reliable measure of the emission from M\,31 at this wavelength. The
foreground Milky Way emission merely contributes to the dispersion of
angles in a given sector and hence to the standard deviation used as
our error estimate.

A further check is applied, by repeating the modelling described in
Sect.~\ref{sec:model} using only the \wwav{6}{11} data. The character
of the deduced regular magnetic field does not substantially change if
the \wav{20} is excluded, though naturally the parameters are less
well defined.

\subsubsection{Polarized intensities}
\label{subsubsec:pi}

We define the average polarized emission of a sector as
$PI=\left(\langle Q\rangle^2+\langle
  U\rangle^2-1.2\sigma_{Q,U}^2\right)^{1/2}$, where $\sigma_{Q,U}$ is
the rms noise in $Q$ and $U$ and provides an approximate correction
for positive bias in $PI$ (Wardle \& Kronberg \cite{Wardle74}). The
$Q$ and $U$ intensities of all pixels in a sector were averaged to
compute $PI$. Errors in non-thermal and polarized intensities were
estimated as the standard deviation between all pixels in a sector as
described in Sect.~\ref{subsubsec:pa}.

In Sect.~\ref{sec:depolar} we compare the degree of polarization at
\wav{20}, where Faraday effects are strong, with that at \wav{6},
where minimal Faraday rotation occurs. It is necessary to smooth the
\wav{20} map to the $3\arcmin$ resolution of the \wav{6} for this
analysis. When smoothed to a resolution of $3\arcmin$ the 'ring' like
polarized emission from M\,31 becomes less distinct than at
$45\arcsec$ and narrow strips of zero polarized intensity become
apparent in the \wav{20} map. These `canals' are interpreted as
depolarization effects in the foreground polarized emission of the
Milky Way by Shukurov \& Berkhuijsen (\cite{Shukurov03}). The
`contamination' of the $3\arcmin$ resolution polarized intensity by
Milky Way emission is therefore probably more serious than for the
polarization angles. The azimuthal pattern of the degree of
polarization at \wav{11} is somewhat similar to that at \wav{20}.
However, the difference in the degrees of polarization at \wav{11} and
\wav{6} is not large enough to allow detailed modelling as a check on
our results in Sect.~\ref{sec:depolar}. Therefore, the observed
\wav{20} polarized intensities are an upper limit on the emission from
M\,31.

\section{The non-thermal disk}
\label{sec:nthdisk}
\begin{table*}
\caption[ ]{Properties of the synchrotron disk in M\,31}
\label{tab:syndisk}
\begin{tabular}{llllllll}
\noalign{\smallskip} \hline \noalign{\smallskip}
$r$  &$I_6$  &$P_6$  &$\Bt$  &$\Br$
  &$\bt$  &$h_6$  &$h_{20}$ \\
(kpc)  &(mJy/beam)  &(\%)  &($\mu$G)  &($\mu$G)  &($\mu$G)  &(pc)
&(pc)\\
\noalign{\smallskip} \hline \noalign{\smallskip}
\pheins 6--\pheins 8  &3.72$\pm$0.06  & 30$\pm$1  &7.3
  &4.9  &5.4  &220  &290\\
\pheins 8--10         &4.71$\pm$0.05  &33$\pm$1          &7.5
  &5.2  &5.4  &240  &330\\
10--12  &4.19$\pm$0.05  &31$\pm$1  &7.1  &4.9  &5.2  &270  &360\\
12--14  &2.71$\pm$0.05  &35$\pm$1  &6.3  &4.6  &4.3  &290  &390\\
\noalign{\smallskip} \hline
\end{tabular}\vfill
\vspace{0.4cm}
Notes: $r$ is the radial range, $I_6$ the average non-thermal
radio intensity per beam area at \wav{6} (HPBW=$3\arcmin$),
$P_6$ the non-thermal degree of polarization at \wav{6} obtained from
the average polarized intensity per beam area divided by $I_6$; $h_6$
and $h_{20}$ the exponential scale heights of the synchrotron
emission at \wav{6} and \wav{20}, respectively;
$\Bt$, $\Br$ and $\bt$ are the average equipartition strengths of
the total, regular and turbulent fields, respectively. 
The uncertainty in the derived field strengths is about 20\%.
See Sect.~\ref{sec:nthdisk} for further explanations.
\end{table*}

Our analysis of depolarization in Sect.~\ref{sec:depolar} requires an
estimate of the scale height of the non-thermal disk and the
discussion of the regular magnetic field, revealed by our model in
Sect.~\ref{sec:model}, is aided by an estimate of the magnetic field
strengths based on equipartition arguments. In this section we derive
both of these quantities.

\subsection{The scale height of the non-thermal emission}\label{SHNTE}
In a study of an arm region in the southwest quadrant of M\,31
Berkhuijsen et al.\ (\cite{Berkhuijsen93}) found that the half-width
of the arms at \wav{20} in the plane of the sky is equal to that of
the total neutral gas (\ion{H}{i}+2H$_2$) suggesting similar scale
heights for radio continuum emission at \wav{20} and neutral gas.

We cannot yet check this for other regions in M\,31, but we can
compare the scale heights at \wav{20} derived by Moss et al.\ 
(\cite{Moss98}) with the scale heights of \ion{H}{i} given by Braun
(\cite{Braun91}). Moss et al.\ (\cite{Moss98}) determined the scale
height of the continuum emission from four cuts parallel to the minor
axis going through the bright `ring' at about $20\arcmin$ on either
side of the centre. The arms were cut at radial distances between 6
and $11\kpc$, and the mean of the exponential scale heights is $325\pm
43\p$. Braun (\cite{Braun91}) described the exponential scale height
of the \ion{H}{i} emission as $h_\mathrm{HI} = (182\pm 37) + (16\pm
3)\, r$, where the radius $r$ is in kpc and $h_\mathrm{HI}$ in pc.
For the same positions as the radio continuum cuts, the mean scale
height of \ion{H}{i} is $310\pm 45\p$. Hence, the scale height of the
radio continuum emission at \wav{20} is the same as that of \ion{H}{i}
within errors. At this wavelength the width of the radio continuum
emission is determined by the synchrotron emission, because the
thermal emission is weak and has a narrower distribution (Berkhuijsen
et al.\ \cite{Berkhuijsen00}). Therefore we take the synchrotron scale
height at \wav{20} equal to the \ion{H}{i} scale height as given by
Braun (\cite{Braun91}).

The synchrotron scale height depends on frequency as $\nu^{-0.25}$, as
observed in NGC\,891 (Hummel et al.\ \cite{Hummel91}) and M\,31
(Berkhuijsen et al.\ \cite{Berkhuijsen91}), thus the scale heights at
\wav{11} and \wav{6} are somewhat smaller than at
\wav{20} (see Table~\ref{tab:syndisk}).

\subsection{The equipartition magnetic field strength}

The transverse component of the total field strength $\Btpe$ (the
quadratic sum of the regular and turbulent components) can be
evaluated from the intensity of the non-thermal emission assuming, for
example, equipartition between the energy densities of magnetic field
and cosmic rays (see Pacholczyk \cite{Pacholczyk70}, Longair
\cite{Longair94}). However, we use a fixed integration interval in
cosmic-ray energy rather than a fixed interval in radio frequency (see
Beck et al.\ \cite{Beck96}). In this case, and for a non-thermal
spectral index $\alpha_\mathrm{n} \simeq 1$, the equipartition field
strength is identical to the minimum-energy field strength. The
polarized intensity yields the strength of the transverse regular
field, $\Brpe$; the transverse turbulent field strength $\btpe$ is
then found from $\btpe^2 = \Btpe^2 - \Brpe^2$. The values of $\Bt$ and
the regular field strength $\Br$ one obtains by deprojection assuming
that $\Br$ is oriented parallel to the plane of M\,31, and $\bt =
\left(\frac{3}{2} \btpe^2\right)^{1/2}$ assuming statistical isotropy.
As Faraday effects are small at \wav{6}, we evaluated the field
strengths from the \wav{6} data.

In Table~\ref{tab:syndisk} we show the average equipartition field
strengths in four $2\kpc$-wide rings covering the bright emission from
M\,31 between $6\kpc$ and $14\kpc$ radius. We used a non-thermal
spectral index $\alpha_\mathrm{n} = 1$ (Berkhuijsen et al.\ 
\cite{Berkhuijsen03}), and the standard ratio of relativistic proton
to electron energy density $k=100$. The line of sight through the
emission layer was taken as $L = 2h_6/\cos i$ with $i=78\degr$; we
note again that the synchrotron scale height depends on $\lambda$.

The magnetic field strengths derived only weakly depend on the errors
in $I$, $L$ and $k$ (as the power $1/(3+\alpha_\mathrm{n})\simeq1/4$).
The main uncertainties are in $L$ and $k$ (about 50\% each) so that
the uncertainty in the derived field strengths in
Table~\ref{tab:syndisk} is about 20\%.

In each ring we also calculated the average magnetic field strength in
the sectors described in Sect.~\ref{subsec:averaging}; within the
errors $\Bt$, $\Br$ and $\bt$ are constant in azimuth.

\section{Overview of the method}
\label{sec:overview}
A short overview of the method we use may help the reader follow the
main part of the paper. We develop two \emph{linked} models in the
following two sections. First, an analysis of the average polarization
angles is used to deduce the underlying structure of the regular
magnetic field in M\,31. One of the parameters in this model,
$\xi_{\lambda}$ in Eq.~(\ref{eq:xi}), can be estimated from a second
model of the Faraday depolarization. However the second model, of the
depolarization, uses rotation measures derived in the first model. We
will try to find solutions that satisfy both models and are consistent
with each other i.e.\ the parameter $\xi_{\lambda}$ is the same,
within errors, in each model.

\section{The 3D structure of the regular magnetic field}
\label{sec:model}
In this section, we deduce the regular magnetic field in M\,31 from
polarization angles of synchrotron emission at \wwwav{6}{11}{20}. The
method used is an extension of that employed by Berkhuijsen et al.\ 
(\cite{Berkhuijsen97}) and is only briefly described here.

\subsection{The model}

The polarization angle of synchrotron emission is given by
\begin{equation}
  \label{eq:psi}
  \psi = \psi_\mathrm{0}(\vec{\Br}) +
  \lambda^2\mathrm{RM_{i}}(\vec{\Br}) + \lambda^2\RMfg,
\end{equation}
where $\psi_\mathrm{0}$ is the intrinsic polarization angle,
$\mathrm{RM_{i}}$ is the Faraday rotation measure in the galaxy,
$\RMfg$ is the Faraday rotation in the Milky Way and $\lambda$ is the
wavelength.

The cylindrical components of $\vec{\Br}=(\Bur,\But,\Buz)$ are
expanded in Fourier series in the azimuthal angle $\theta$,
\begin{eqnarray}
  \label{eq:B}
  \Bur & = & B_0\sin p_0 + B_1\sin p_1\cos(\theta-\beta_1) \nonumber \\
      & & +\, B_2\sin p_2\cos2(\theta-\beta_2),\nonumber \\
  \But & = & B_0\cos p_0 + B_1\cos p_1\cos(\theta-\beta_1) \\
      & & +\, B_2\cos p_2\cos2(\theta-\beta_2),\nonumber\\
  \Buz & = & B_{z0} + B_{z1}\cos(\theta-\beta_{z1}) + B_{z2}
  \cos2(\theta-\beta_{z2}),\nonumber
\end{eqnarray}
where $B_m$ and $B_{zm}$ are the amplitude of the mode with azimuthal
wave number $m$ in the horizontal and vertical fields, $p_m$ is the
pitch angle of the $m$'th horizontal Fourier mode (i.e.\ the angle
between the field and the local circumference) and $\beta_m$ and
$\beta_{zm}$ are the azimuths where the non-axisymmetric modes are
maximum. Expressions for $\psi_\mathrm{0}$ and RM in terms of the
expansions shown in Eq.~(\ref{eq:B}) are given in Eqs.\,(A3) and (6)
of Berkhuijsen et al.\ (\cite{Berkhuijsen97}). We note here that
$\psi_\mathrm{0}$ depends on magnetic field components in the sky
plane whereas RM depends on those along the line of sight.  Therefore,
fitting Eq.~(\ref{eq:psi}) allows us to obtain all three components of
$\vec\Br$.  Since the observed polarization angle depends on RM, i.e.\ 
on the product of the magnetic field strength, thermal electron
density and the path length, the amplitudes of the Fourier modes are
obtained from fitting in terms of the variables $R_m$ whose dimension
is $\FRM$:
\begin{equation}
  \label{eq:FRM}
  R_m = 0.81 \left( \frac{B_m}{1\mkG} \right)
        \left( \frac{\langle n_\mathrm{e} \rangle}{1\cmcube} \right)
        \left( \frac{L}{1 \p} \right),
\end{equation}
where $\langle n_\mathrm{e} \rangle$ and $L$ are the average density
of thermal electrons  and the line of sight path length through the thermal
disk in a given ring.

Only a fraction of the synchrotron emitting disk may be visible at a
given wavelength due to Faraday depolarization.  Therefore,
observations of polarized emission at different wavelengths probe the
galactic disk to different depths and our analysis can reveal
variations in the disk parameters along the line of sight.  This has
allowed Berkhuijsen et al.\ \cite{Berkhuijsen97}) to reveal a
two-component magneto-ionic structure in M\,51 comprising a disk and a
halo. M\,31 does not have an extensive synchrotron halo (Gr\"ave et
al.\ \cite{Graeve81}), and so we consider one-component (i.e.\ disk
only) fits where the galactic disk is probed to different depths at
different wavelengths.  Correspondingly, Faraday rotation is scaled by
a wavelength dependent factor $\xi_{\lambda}\leq 1$, so
\begin{equation}
  \label{eq:xi}
  \mathrm{RM_{i}=\xi_{\lambda} RM\disk},
\end{equation}
where $\mathrm{RM\disk}$ is the Faraday rotation measure produced
through the whole disk thickness (observable at short wavelengths),
and $\xi_\lambda$ can be understood as the fraction of the disk
thickness transparent to polarized emission at a wavelength $\lambda$. In
Sect.~\ref{sec:depolar} we discuss depolarizing mechanisms in detail,
and from models of the observed depolarization we adopt $\xi_6=1.0$,
$\xi_{11}=0.96$ and $\xi_{20}=0.75$ (see Eqs.~\ref{eq:Deltaz} and
\ref{eq:xideltaz}). The fitted parameters describing the magnetic
field were found to be rather insensitive to the adopted values of
$\xi_{\lambda}$.

Using Eqs. (\ref{eq:psi}), (\ref{eq:B}) and (\ref{eq:xi}) we fit the
modelled, three-dimensional $\vec\Br$ to the observed polarization
angles in a ring, simultaneously for all wavelengths, by minimizing
the residual
\begin{equation}
  \label{eq:residual}
  S = \sum_{\lambda,n}\left\lbrack\frac{\psi_{n}
    - \psi(\theta_{n})}{\sigma_{n}}\right\rbrack^2 ,
\end{equation}
where $\psi_{n}$ is the observed angle of polarization,
$\psi(\theta_{n})$ the modelled angle and $\sigma_{n}$ are the
observational errors. The $\chi^2$ test is used to ensure that the
fit, for all wavelengths in a ring, is sufficiently close to the
measured angles. Application of the Fisher test verified that the fits
are equally good at each individual wavelength
(see Berkhuijsen et al.\ \cite{Berkhuijsen97}). This model is aimed at
analysis of a global structure of the magnetic field and is not
devised to capture local details in the field structure.  Therefore,
where a few data points deviate very strongly from the general pattern
they can be discarded to obtain a statistically good fit. The number
of points discarded for this reason was $6\%$ of the total
measurements and most of these points occur where $\psi$ varies very
strongly with $\theta$, leading to underestimated errors. The
exclusion of points is discussed further in
Sect.~\ref{subsec:discard}.

We determine the errors in the fit parameters by varying them
independently and in paired combinations to determine the parameter
ranges consistent with the $\chi^2$ test. For fits requiring a small
number of parameters, we checked these error estimates by plotting
contours of the residual $S$ in the parameter space. The resulting errors,
quoted below, are all $2\sigma$ deviations.

\subsection{Results of fitting}

\begin{figure}
  \includegraphics[width=0.45\textwidth]{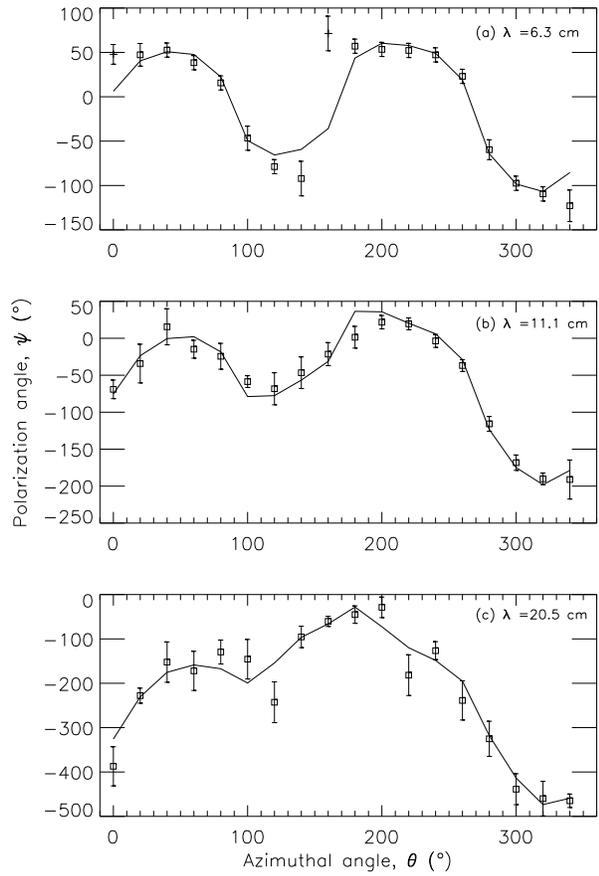}
  \caption{Polarization angles
    ($\psi$, measured from the local radial direction in the plane of
    M\,31) against azimuth ($\theta$) for the ring 6--8$\kpc$. Fit
    (solid) and observations (squares with error bars, horizontal
    lines with error bars show excluded points) are shown for \wav{6},
    \wav{11} and \wav{20.5} from top to bottom. The error bars show
the $1\sigma$ deviations.}
    \label{fig:fitRing1}
\end{figure}
\begin{figure}
  \includegraphics[width=0.45\textwidth]{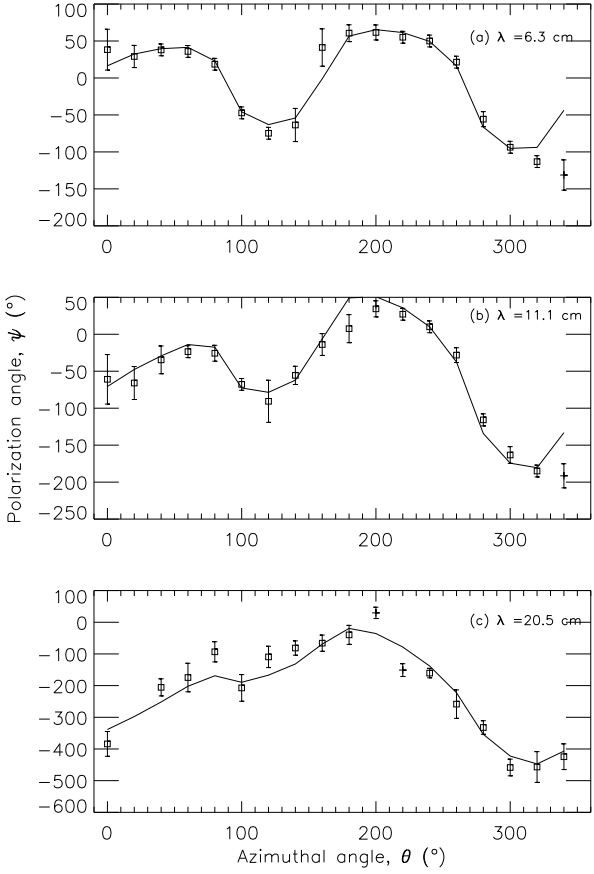}
  \caption{As in Fig.~\ref{fig:fitRing1} but
    for the ring 8--10$\kpc$.}
  \label{fig:fitRing2}
\end{figure}
\begin{figure}
  \includegraphics[width=0.45\textwidth]{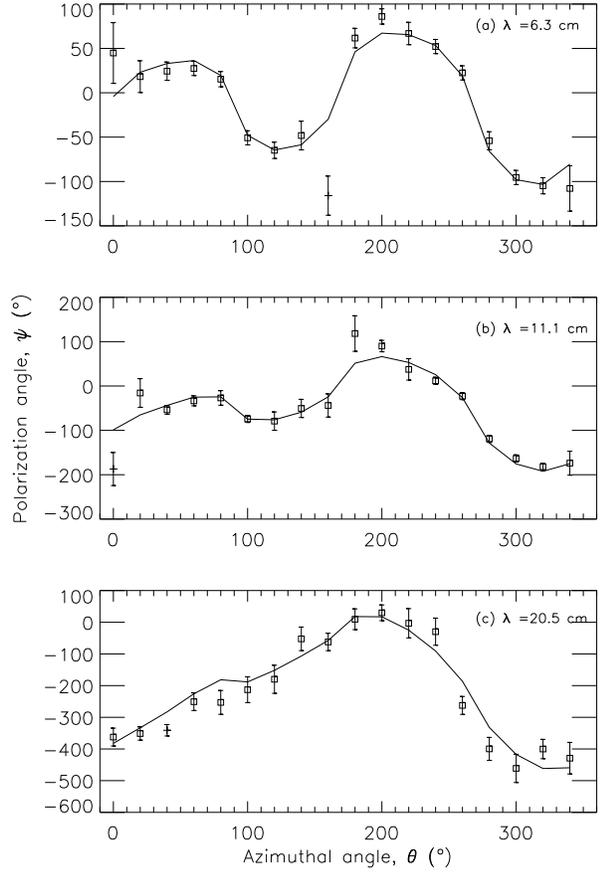}
  \caption{As in Fig.~\ref{fig:fitRing1} but
    for the ring 10--12$\kpc$.}
  \label{fig:fitRing3}
\end{figure}
\begin{figure}
  \includegraphics[width=0.45\textwidth]{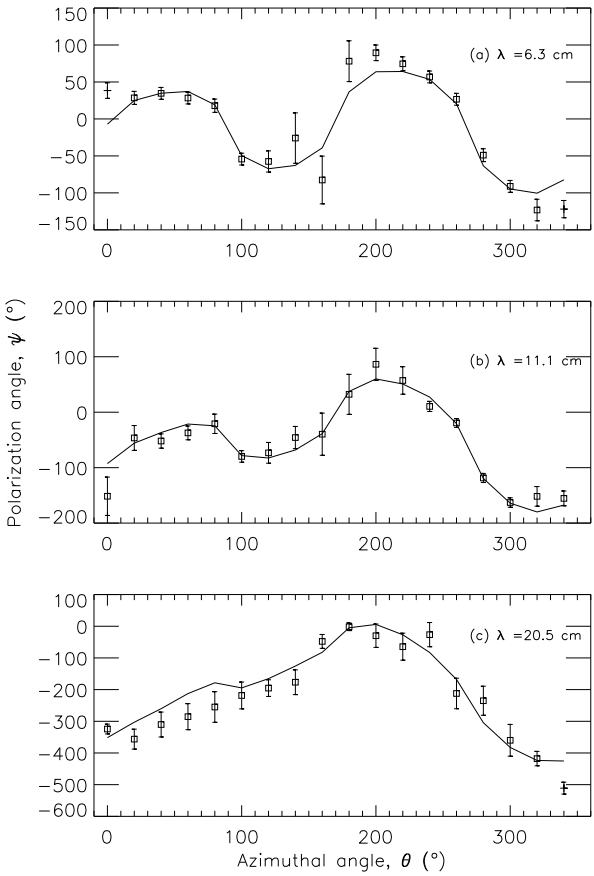}
  \caption{As in Fig.~\ref{fig:fitRing1} but
    for the ring 12--14$\kpc$.}
  \label{fig:fitRing4}
\end{figure}
\begin{table}
\caption{Parameters of the fitted model and their $2\sigma$ errors.
$RM_\mathrm{fg}$ is the Faraday rotation measure arising in the Milky Way,
$R_m$ and $p_m$ are the amplitude and pitch angle of the mode with wave number
$m$, and $\beta_m$ is the azimuth where a mode with azimuthal wave number $m$
is maximum. The minimum value of the residual and the value of $\chi^2$ are
shown  for each fit in the bottom lines.}
\begin{center}
\begin{tabular}{llcccc} \hline \noalign{\smallskip}
  & Units & \multicolumn{4}{c}{Radial range (kpc)}  \\
  & & 6--8 & 8--10 & 10--12 & 12--14  \\
\hline \noalign{\medskip} RM$_{\rm fg}$ & $\rm{rad\,m^{-2}}$ &
  $-93$\,\scriptsize{$\pm5$} & $-99$\,\scriptsize{$\pm5$} &
  $-93$\,\scriptsize{$\pm5$} & $-89$\,\scriptsize{$\pm4$} \\ \noalign{\smallskip}
$R_0$ & $\rm{rad\,m^{-2}}$ &
  $+83$\,\scriptsize{$\pm7$} & $+96$\,\scriptsize{$\pm9$} &
  $+115$\,\scriptsize{$\pm9$} & $+99$\,\scriptsize{$\pm6$} \\ \noalign{\smallskip}
$p_0$ & deg &
  $-13$\,\scriptsize{$\pm4$} & $-19$\,\scriptsize{$\pm3$} &
  $-11$\,\scriptsize{$\pm3$} & $-8$\,\scriptsize{$\pm3$}  \\ \noalign{\smallskip}
$R_2$ &  $\rm{rad\,m^{-2}}$ &
  $+45$\,\scriptsize{$\pm10$} & & & \\ \noalign{\smallskip}
$p_2$ & deg &
  $-2$\,\scriptsize{$\pm12$} & & & \\ \noalign{\smallskip}
$\beta_2$ & deg &
  $-43$\,\scriptsize{$\pm7$} & & & \\ \noalign{\medskip}
\hline \noalign{\medskip} $S$ & & $58$ & $59$ & $62$ & $62$
\\\noalign{\smallskip}
$\chi^2$ & & $63$ & $63$ & $65$ & $65$  \\
\noalign{\medskip}
  \hline
  \label{tab:results}
\end{tabular}
\end{center}
\end{table}

Figures \ref{fig:fitRing1} to \ref{fig:fitRing4} show the variation of
observed polarization angles ($\psi$, measured anti-clockwise from the
local radial direction in the plane of M\,31) with azimuthal angle
$\theta$ and the fits for each ring. The fitted parameters are given
in Table~\ref{tab:results}. Generally we find that an axisymmetric
field, lying parallel to the galactic midplane provides the best fit
to the measured polarization angles.  For the innermost ring a weaker,
$\pi$-periodic ($m=2$) mode is added to the dominant axisymmetric
($m=0$) mode. The $m=2$ mode will produce a $\pi/2$ periodicity in
$\mathrm{RM}$.

The fitted $\mathrm{RM_{fg}}$ is constant, within errors, between
adjacent rings and varies weakly across the whole radial range in
agreement with the expected small fluctuations in foreground RM from
our Galaxy in the direction of M\,31 (Han et al.\  \cite{Han98}). This
is an important reliability check for the model; the values of
$\mathrm{RM_{fg}}$ in Table~\ref{tab:results} were independently
derived for each ring by fitting a non-linear model to the
observational data. It is reassuring that there is agreement between
rings within errors and with earlier estimates. The value of $\RMfg$
is broadly consistent with earlier estimates of $-88 \pm2\FRM$ (Beck
\cite{Beck82}), $-100\pm33\FRM$ (Ruzmaikin et al.\
\cite{Ruzmaikin90}), $-93\pm3\FRM$ (Han et al.\ \cite{Han98}) and
$-92\pm3\FRM$ (Berkhuijsen et al.\ \cite{Berkhuijsen03}).

The median amplitude of the axisymmetric mode $R_0$ reaches a maximum at
$R\simeq11\kpc$, the radius of the well known bright radio 'ring' of M\,31.
However, the maximum is only marginally pronounced, and the values of $R_0$
only show radial variation at the $2\sigma$ level. This implies that the
synchrotron ring in M\,31 is prominent either because the synchrotron
emissivity depends on a high power of the magnetic field strength or because
the density of relativistic electrons is higher in the ring. The underlying
maximum in the magnetic field itself is very weak, of about 20\%, or even less
if the thermal electron density has a maximum in the ring.

We now describe the fits for each ring in detail.

\subsubsection{The ring $r=\mbox{}$6--8\,kpc}
\label{subsec:model:ring1}
A combination of a strong $m=0$ mode perturbed by a weaker $m=2$ mode
provides a good fit for the innermost ring (i.e., one that satisfies
both the $\chi^2$ and Fisher tests). The rotation measure in this ring
varies by a factor of 3 between the maxima ($\sim130 \FRM$ at
$\theta=130\degr,\,310\degr$) and minima ($\sim 45 \FRM$ at
$\theta=50\degr,\, 230\degr$). If $\langle n_\mathrm{e} \rangle L$ is
about constant in the ring, the field strength varies by the same
factor. The pitch angle of the $m=2$ mode is small but leads to a
variation of $\pm 10\degr$ in the mean pitch angle of the regular
magnetic field, $p=\arctan(\Bur/\But)$, with minimum pitch angles of
$p=-25\degr$ and maxima of $p=-9\degr$ at $\theta=40,220\degr$ and
$140,320\degr$, respectively.

To achieve this fit we excluded two data points (at \wav{6} the sectors
$\theta=0\degr$ and $160\degr$) out of 54. Both sectors are in the region of a
vary rapid change in $\psi$, so it is plausible that the error in $\psi$ is
underestimated in the two sectors. If we try to obtain a good fit for the
combination $m=0+1$, it is necessary to exclude four measurements (at \wav{6}
$\theta=0\degr,\,160\degr,\,340\degr$ and at \wav{11} $\theta=180\degr$) and
the fitted $\RMfg=-113\pm 2\FRM$ is not consistent with $\RMfg$ in the other
rings. Nine measurements must be excluded in order to achieve a fit using only
the axisymmetric $m=0$ mode, so the addition of three extra parameters
describing the $m=2$ mode is supported by the use of seven extra data points.

A possible explanation for the $m=2$ mode in this ring can be that the
disk inclination angle $i$ is different from that in the other rings.
Braun (\cite{Braun91}) argues that the inclination angle of the
\ion{H}{i} disk varies significantly along radius in M\,31.

Another, more plausible possibility is that the $m=2$ component is a response
to the two armed spiral pattern, but restricted to the thin magneto-ionic
disk. The latter restriction is needed to explain why this magnetic field
model does not deliver a good fit to the Faraday depolarization in this ring
discussed in Sect.~\ref{subsec:depolar:model}. As we argue there, the
depolarization, due to a Faraday screen, occurs in the upper layers where the
field is basically axisymmetric

\subsubsection{The rings 8--10, 10--12 and 12--14 kpc}
A satisfactory fit using only the axisymmetric $m=0$ mode is found for each of
these rings. The mode amplitude reaches a weak maximum in the ring $10-12\kpc$
and then decreases in the outermost ring. The pitch angle of the regular
magnetic field becomes smaller (i.e., the field becomes more tightly wound
with increasing radius (see Sect.~\ref{subsec:pa}).

The fit for the ring $8-10\kpc$ requires the omission of four measured
$\psi$ out of the total of 54, two near the major axis at
$\theta=340\degr$ for \wwav{6}{11}, and two at \wav{20},
$\theta=200\degr$, $220\degr$. For the ring 10--12\,kpc three
measurements must be omitted to achieve the $m=0$ fit, two on the
major axis (at \wav{6}, $\theta=160\degr$ and \wav{11},
$\theta=0\degr$), along with the sector $\theta=40\degr$ at \wav{20}.
Finally, the measurements at \wav{6} $\theta=0\degr, 340\degr$ and
\wav{20} $\theta=340\degr$ are omitted in the outermost ring.

\begin{figure}
  \includegraphics[width=0.35\textwidth,angle=0]{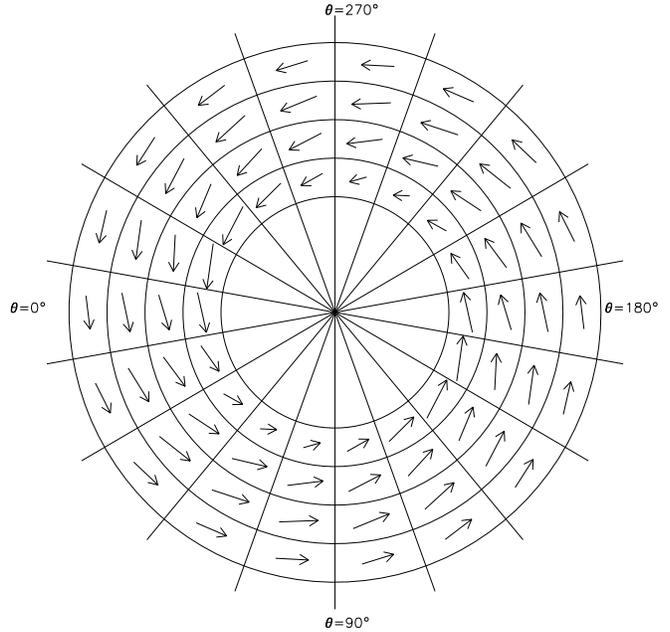}
  \caption{Face-on view of M\,31 showing sectors and regular magnetic
    field vectors obtained from the fits shown in
    Table~\ref{tab:results}. The grid radii are 6, 8, 10, 12 and
    14$\,\rm{kpc}$. The length of the vectors is proportional to
    $\vec\Br$.}
\label{fig:polar}
\end{figure}

Figure~\ref{fig:polar} shows a face-on view of the galactic disk with
the sector grid and the fitted regular magnetic field vectors shown in
each sector.  The azimuthal component of the field is stronger than
the radial component in all sectors (that is, the pitch angle is
rather small). The effect of the $\pi$-periodic, $m=2$, mode in the
innermost ring can be clearly seen in the varying length and direction
of the magnetic field vectors.

\subsection{Excluded measurements}
\label{subsec:discard}
In order to include all of the observations in any of the rings, we find that
more than two extra modes must be added to the magnetic field models discussed
above. For example, in the ring 10--12\,kpc we cannot achieve a good fit with
the combination $m=0+1+2$ even though an extra 6 parameters are used to try
and accommodate three previously excluded measurements.  This strongly
suggests that either (i) the excluded sectors are not dominated by any
large-scale structure but by localised perturbations of a more regular
underlying pattern or (ii) the errors in the omitted polarization angles are
underestimated.

All but two of the excluded measurements lie close to the major axis
of M\,31. Here, the detected polarized emission is weakest as the
small pitch angle of the regular magnetic field means that its
component perpendicular to the line of sight, $\Brpe$, is small near the
major axis. Also, near the major axis of a highly inclined galaxy with
a strongly azimuthal regular magnetic field, the observed polarization
angle (in the sky plane) changes rapidly. These effects can lead to
underestimation of the errors in $\psi$ for sectors near the major
axis. Furthermore, any deviation from an axisymmetric field (e.g.\ due
to inter-arm bridges) near the major axis of M\,31 contributes to
the line-of-sight magnetic field and distorts the smooth pattern of Faraday
rotation measures.

\subsection{Magnetic pitch angles}
\label{subsec:pa}
The pitch angles of the regular magnetic field are $p\simeq -17\degr$
between $6<r<10\kpc$ and then become smaller with increasing radius,
reaching $p\simeq -8\degr$ in the ring $12<r<14\kpc$. These values are
more reliable than earlier estimates -- more data are used in the
modelling and interpretation methods have improved -- but are in broad
agreement with the results of Beck (\cite{Beck82}), Ruzmaikin et al.\ 
(\cite{Ruzmaikin90}) and Berkhuijsen et al.\ (\cite{Berkhuijsen03}).
The regular magnetic fields maintained by galactic dynamo action must
have a non-zero pitch angle, since the dynamo generates both radial
and azimuthal magnetic field components (Shukurov\ \cite{Shukurov00}).
The sign, magnitude and radial trend of the magnetic field pitch
angles are in broad agreement with the predictions of a range of
dynamo models for M\,31 (\cite{Shukurov00}).

Observations of CO (Gu{\'e}lin et al.\ \cite{Guelin00}) and \ion{H}{i}
(Braun \cite{Braun91}) have been fitted with logarithmic spirals
tracing the gaseous arms with a constant pitch angle of $\simeq
-7\degr$. In those nearby spiral galaxies where density waves are
believed to be present, the regular magnetic fields generally follow
the spiral structure (see Beck (\cite{Beck96}) and references
therein). The difference between the magnetic and spiral arm pitch
angles for $6<r<12\kpc$ may be because density waves are absent or
very weak in M\,31. A detailed comparison with the spiral structure,
seen e.g.\ in the CO line emission, is required to clarify the
relation between the magnetic and gas spirals.

\section{Depolarization}
\label{sec:depolar}

The observed degree of polarization of non-thermal emission from
external galaxies is generally less than the intrinsic maximum of
$P_{0}\simeq 0.75$ for a completely regular magnetic field structure.
The reduction in the degree of polarization
can be due to the physical properties of the ISM in the
galaxy and to effects arising from the finite size of the telescope
beam. By investigating depolarization mechanisms we can recover
information about the ISM.

A convenient measure of depolarization is the ratio of relative polarized
intensities at two wavelength, i.e.,
\begin{equation}
  \label{eq:DP}
  \DP_{\lambda_1/\lambda_2}=\frac{P(\lambda_1)}{P(\lambda_2)},
\end{equation}
where $P(\lambda)$ is the degree of polarization at a wavelength
$\lambda$; $\DP_{\lambda_1/\lambda_2}=1$ means no depolarization
between the two wavelengths. A variety of depolarization mechanisms in
radio sources are discussed by, e.g., Burn (\cite{Burn66}), Pacholczyk
(\cite{Pacholczyk70}) and, specifically for spiral galaxies, by
Sokoloff et al.\ (\cite{Sokoloff98}).  A description of several
concurrent depolarization mechanisms can be rather complicated.
Wavelength-independent depolarization and that due to Faraday rotation
(and so wavelength-dependent) can be easily isolated as they result in
independent factors in the total depolarization; in Eq.~(\ref{eq:DP})
the wavelength independent contributions to depolarization at $\lambda_1$ and
$\lambda_2$ are equal to each other and cancel, so that the observed $\DP$ is a
measure of depolarization due to Faraday effects. Among wavelength
dependent depolarization mechanisms, depolarization in a Faraday
screen and in the synchrotron source can be disentangled because they
occur in non-overlapping regions.  However, distinct Faraday
depolarization mechanisms that occur within the same volume cannot be
represented as independent factors in the total depolarization in the
general case.  An approximation that allows one to separate the
internal Faraday dispersion from other depolarizing effects in the
synchrotron source has been suggested by Sokoloff et al.\
(\cite{Sokoloff98}, their Sect.\ 6.3) and called the `opaque layer
approximation'. In Appendix~\ref{app} we briefly describe the
different depolarization mechanisms affecting observations of external
galaxies and give the equations used later in this section to model
the observed depolarization.

Depolarization of the non-thermal emission must be carefully considered when
interpreting the data. For example, the synchrotron disk can be transparent to
polarized emission at short wavelengths but opaque at longer wavelengths (see
e.g.\ Berkhuijsen et al.\ \cite{Berkhuijsen97}). Therefore, the amount of
Faraday rotation is no longer proportional to $\lambda^2$. (This is the
motivation for introducing the parameter $\xi_{\lambda}$ in
Sect.~\ref{sec:model}.) First though, we look at the observed depolarization
in a qualitative way.  Then we attempt to construct a model for the
observations, in terms of parameters describing the state of the ISM.

\subsection{The dominant depolarization mechanism in M\,31}

\begin{figure}
  \includegraphics[width=0.45\textwidth]{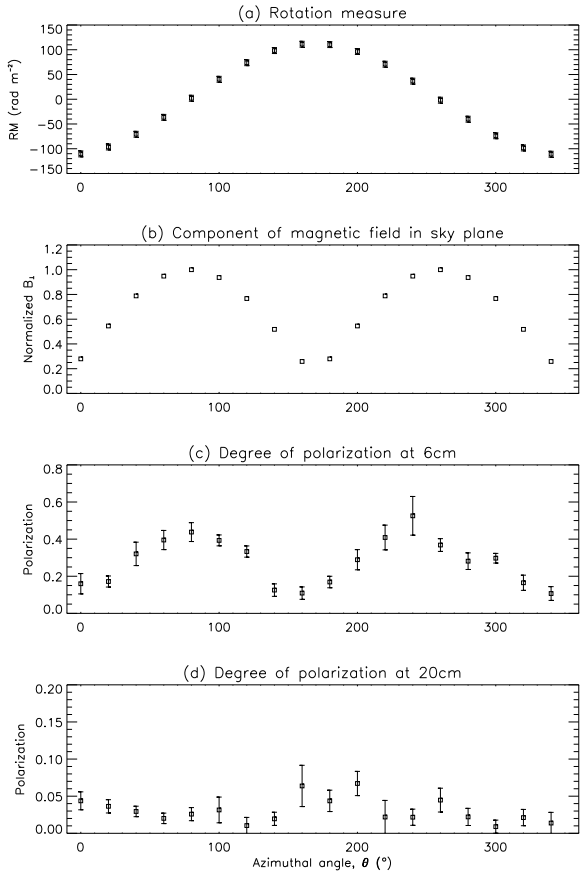}
  \caption{For the ring $10<r<12\kpc$. (a) Rotation measures derived from
    the fitted magnetic field shown in Table~\ref{tab:results}. (b)
    The normalized amplitude of the component of the regular magnetic
    field lying in the sky plane, $\Brpe$, derived from the fitted
    magnetic field shown in Table~\ref{tab:results}. (c) The observed
    degree of polarization ($P_6$=polarized intensity/total nonthermal
    intensity) at \wav{6}. (d) The observed degree of polarization at
    \wav{20} ($P_{20}$).}
  \label{fig:RM}
\end{figure}

The ring $10<r<12\kpc$ is chosen for an initial, closer look at
depolarization. In Fig.~\ref{fig:RM} we show the azimuthal variation
of some key properties in this ring. $RM$ and $\Brpe$ have been
derived from the polarization angle model presented in
Sect.~\ref{sec:model}; $\Brpe$ has been obtained assuming that
$\langle n_\mathrm{e} \rangle L$ (see Eq.~3) is constant in azimuth,
and normalized. We also show the observed degrees of polarized
emission at \wav{6} and \wav{20} ($P_{6}$ and $P_{20}$ respectively).
The pattern of $RM$ versus azimuthal angle is determined by the
geometrical variation of $\Brpa$, with the strongest $RM$ near the
major axis where the regular magnetic field lies along the line of
sight to M\,31. Note also that the sine-like variation of $RM$ results
in the strongest gradients in $RM$ lying near the minor axis.
Furthermore, Berkhuijsen et al.~(\cite{Berkhuijsen03}) noted that the
azimuthal variation of the polarized emission at \wav{6} is almost
completely due to the geometrical variation of $\Brpe$ with azimuthal
angle.  Figures~\ref{fig:RM}b and \ref{fig:RM}c clearly show that this
also holds for $P_{6}$, with the $P_{6}$ highest near the minor axis
where $\Brpe$ is strongest. In contrast, the degree of polarization at
\wav{20}, $P_{20}$, has a less marked azimuthal variation. If
anything, the pattern of $P_{20}$ is the inverse of $P_6$, but with a
lower amplitude. The wavelength dependent depolarization,
$\DP_{20/6}$, is obtained by dividing $P_{20}$ (Fig.~\ref{fig:RM}d) by
$P_6$ (Fig.~\ref{fig:RM}c).

\begin{figure}
  \includegraphics[width=0.45\textwidth]{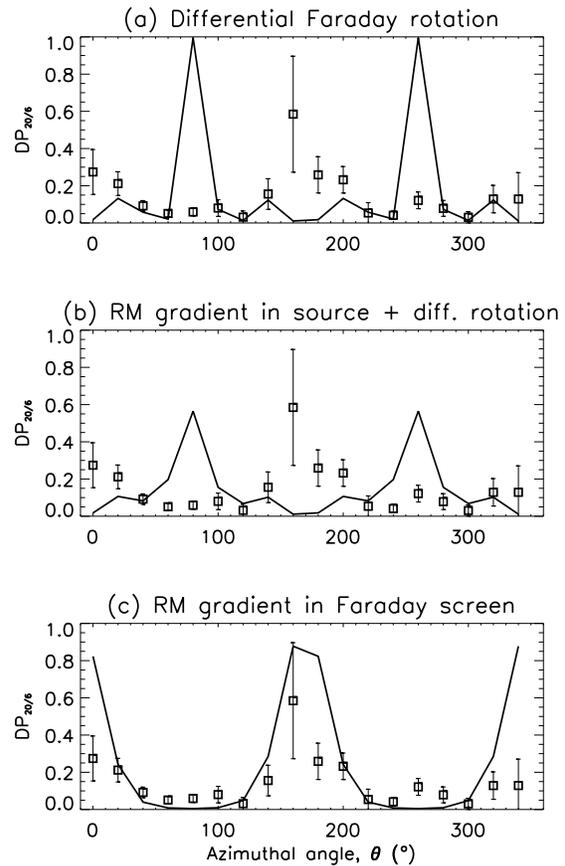}
  \caption{Observed (squares with error bars) and expected (solid line)
    depolarization between \wav{20} and \wav{6}, $\DP_{20/6}$, for the
    ring $10<r<12 \kpc$ assuming various depolarization mechanisms.
    The azimuthal angle $\theta$ is measured counterclockwise from the
    northern major axis.  Depolarization due to
{\bf(a)} differential Faraday rotation using Eq.~(\ref{eq:DPreg}),
{\bf(b)} RM gradients in the
    synchrotron source and differential Faraday rotation from
    Eq.~(\ref{eq:DPgradint}) and
{\bf(c)} RM gradients in a Faraday
    screen given by Eq.~(\ref{eq:DPgradext}) are represented
    by solid lines.  RM and the gradient in RM are derived from the
    fitted magnetic field described in Sect.~\ref{sec:model}. We
    conclude that foreground RM gradients, illustrated in panel (c),
dominate other wavelength
    dependent depolarization mechanisms.}
  \label{fig:DPMechanism}
\end{figure}

Can we recognize the signature of any of the depolarization mechanisms
discussed in Appendix~\ref{app} in the observations? Before
considering a model based on a combination of effects it
is instructive to consider each of these mechanisms separately.

The effect of wavelength independent depolarization is removed by
considering Eq.~(\ref{eq:DP}). By comparing the ratio of the observed
degrees of polarization at \wav{20} and \wav{6}, $\DP_{20/6}=P_{20}/P_{6}$,
with that expected from Eqs.~(\ref{eq:DPreg}) to (\ref{eq:DPgradext}),
we identify which wavelength dependent depolarization mechanisms are
dominant.

The observed $\DP_{20/6}$, plotted in Fig.~\ref{fig:DPMechanism}, has
a marked azimuthal variation with strong depolarization at \wav{20}
near the minor axis ($\theta = 90\degr$ and $270\degr$, where
$\DP_{20/6}\simeq0.1$) and less depolarization on the major axis
($\theta = 0\degr$ and $180\degr$ where $\DP_{20/6}\geq 0.4$). Note
that the observed $\DP_{20/6}(\theta)$ is roughly proportional to the
derivative of RM$(\theta)$ (Fig.~\ref{fig:RM}).

Gradients in the foreground RM due to magnetic fields in the Milky
Way, $\RMfg$, in the direction of M\,31, are weak (Han et
al.\ \cite{Han98}) and unlikely to cause the observed variation of RM
and DP with azimuth in M\,31. Thus, depolarization must occur within
M\,31. For the rest of the analysis of depolarization we consider the
RM intrinsic to M\,31, $\RMi = \RM - \RMfg$.

The smooth, sinusoidal azimuthal variation of RM (Fig.~\ref{fig:RM}a)
can be completely accounted for by an azimuthal variation of $\Brpa$
deduced in Sect.~\ref{sec:model}, indicating that $\langle
n_\mathrm{e}\rangle L$ is indeed roughly constant in azimuth.  The turbulent
magnetic field, $\bt$, derived using the equipartition approach
described in Sect.~\ref{sec:nthdisk}, is also constant in azimuth for
each ring. Therefore the dispersion in RM, $\sigma_\mathrm{RM}$, and
hence depolarization due to Faraday dispersion, is roughly constant at
a given radius, and the azimuthal variation in $\DP_{20/6}$ cannot be
explained by Faraday dispersion [Eqs.~(\ref{eq:DPin}) and
(\ref{eq:DPex})]. This does not mean that Faraday dispersion is
ineffective in M\,31, but rather that the strong azimuthal pattern in
$\DP_{20/6}$ cannot be explained by this mechanism.

The remaining wavelength dependent depolarizing mechanisms are all
caused by the regular magnetic field: differential Faraday rotation,
RM gradients within the emitting layer, and RM gradients in a
foreground Faraday screen. The first two effects are unavoidable while
the third effect requires the existence of a 'thick disk' of magnetic
fields and thermal gas invisible in synchrotron emission. Differential
Faraday rotation in the source is strongest near the major axis where
the line-of-sight magnetic field $\Brpa$ is maximum, resulting in a
depolarization pattern very different from that observed
(Fig.~\ref{fig:DPMechanism}a). The azimuthal gradient in $\Brpa$ is
also maximum near the minor axis (where it changes sign).  Therefore,
depolarization due to gradients in RM in the synchrotron source is
strong near the minor axis and weak near the major axis, but still
does not overcome the differential Faraday rotation that produces a
different pattern (Fig.~\ref{fig:DPMechanism}b). On the other hand a
foreground Faraday screen does not produce any differential Faraday
rotation, and so depolarization due to the RM gradients in a
foreground screen is dominant, producing a correct pattern shown in
Fig.~\ref{fig:DPMechanism}c.

Thus, the global pattern of the azimuthal variation of $\DP_{20/6}$
can only be reproduced by depolarization due to RM gradients in a
Faraday screen (the bottom frame of Fig.~\ref{fig:DPMechanism}). This
mechanism must be the dominant cause of the azimuthal pattern in
wavelength dependent depolarization. This is true in the whole radial
range $6\leq r\leq14\kpc$. Berkhuijsen et al.\ (\cite{Berkhuijsen03})
found that contours of RM and $\DP_{11/6}$ are often perpendicular to
each other where they cross (see their Fig.~14) and noted that this
suggests RM gradients as an important cause of depolarization.
Earlier, Berkhuijsen \& Beck (\cite{Berkhuijsen90}) found that RM
gradients were primarily responsible for depolarization in the
southwestern quadrant of M\,31 and Horellou et al.\ 
(\cite{Horellou92}) observed that contours of $\DP$ and RM are
perpendicular at crossing points for the galaxy M\,51.

The minima in $\DP_{20/6}$ produced by the Faraday screen are
noticeably deeper than those observed (at $\theta\approx90\degr$ and
$270\degr$ in Fig.~\ref{fig:DPMechanism}c). As discussed in
Sect.~\ref{subsec:depolar:model}, this can be explained by other, less
important depolarization mechanisms.

\subsection{The thermal and synchrotron disk scale heights}
\label{subsec:depolar:model}

\begin{figure}
  \includegraphics[width=0.45\textwidth]{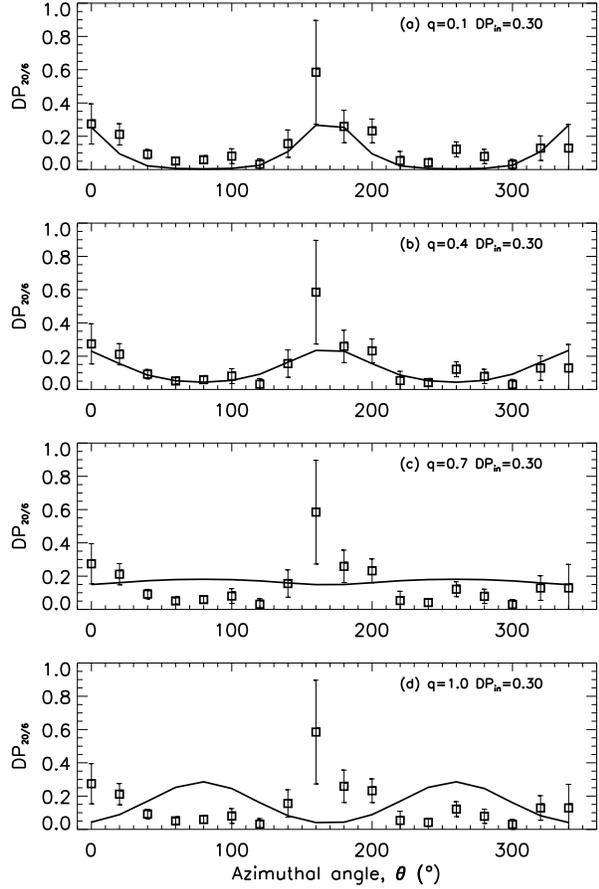}
  \caption{Observed (squares with error bars) and expected (solid line)
    depolarization between \wav{20} and \wav{6}, $\DP_{20/6}$, due to
    RM gradients and differential Faraday rotation using
    Eqs.~(\ref{eq:DPtot}), (\ref{eq:RMdisk}) and (\ref{eq:RMscreen}),
    for the ring $10<r<12 \kpc$ and with different ratios of the
    synchrotron to thermal disk scale heights, $q=h\syn/h\therm$.}
  \label{fig:DPRing3}
\end{figure}

We have identified RM gradients in a Faraday screen as the dominant
depolarizing mechanism responsible for the observed azimuthal pattern
of $\DP_{20/6}$ in M\,31. The fit to observations in
Fig.~\ref{fig:DPMechanism}c can be improved by including other
depolarizing effects, especially Faraday dispersion. Also, the
effectiveness of the Faraday screen depends upon its relative
thickness, compared to that of the synchrotron emitting layer. Now we
attempt to recover information about the relative heights of the
emitting and Faraday rotating layers from fitting the depolarization.

\begin{figure}
  \includegraphics[width=0.3\textwidth,keepaspectratio,angle=0]{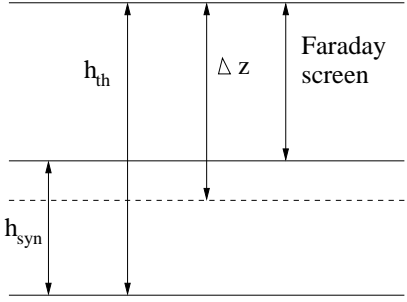}
  \caption{Sketch showing the scale heights of the thermal disk,
    $h\therm$, synchrotron disk, $h\syn$, and the depth in the thermal
    disk from which polarized emission is visible at \wav{20}, $\Delta
    z$. The galactic midplane is at $z=0$, i.e., at the bottom of the
    figure.  The deeper layers of the synchrotron disk are invisible
    in polarized emission because of internal Faraday dispersion. This
    sketch is only an illustration: note that $\Delta z$ can exceed
    $h\therm$ if more than half the disk thickness is visible.
    However, $h\therm-h\syn\leq\Delta z\leq h\therm+h\syn$ with the
    extreme values corresponding to $\DP_{20/6}=0$ and 1,
    respectively.}
    \label{fig:diskfig}
\end{figure}

A full description of depolarization due to the regular magnetic field
(i.e., differential Faraday rotation, RM gradients inside the
synchrotron emitting layer and in a Faraday screen) is given by the
product of Eqs.~(\ref{eq:DPgradint}) and (\ref{eq:DPgradext}).  The
intrinsic Faraday rotation measure $\RMi$ and its increment
$\Delta\RM$ across each sector can be calculated from the fits for the
regular magnetic field discussed in Sect.~\ref{sec:model}. We split
$\RMi$ into two components, $\RM\disk$ arising within the synchrotron
disk and $\RM\screen$ arising in the part of the thermal layer above
the synchrotron disk (see Fig.~\ref{fig:diskfig}); the scale height of
the synchrotron layer is taken from Sect.~\ref{SHNTE}. The first
component will produce depolarization due to differential Faraday
rotation, but the latter will only contribute to Faraday screen
effects. The gradient in $\RMi$ is similarly split into
$\Delta\RM\disk$ and $\Delta\RM\screen$.  In terms of these variables,
the degree of polarization with allowance for Faraday dispersion,
differential Faraday rotation and rotation measure gradients in both
the thermal disk and Faraday screen is given by

\[
  P(\lambda) =  P_0 P_\mathrm{in}(\lambda)P_\mathrm{ex}(\lambda)
  \left| \exp\left[2i{\RM_0}\screen\lambda^2 \! - \!
  2\left(\Delta\RM\screen\lambda^2\right)^2\right]
\phantom{int_0^1}
\right.
\]
\begin{equation}
 \label{eq:DPtot}
\left. \mbox{}  \qquad\quad  \times\int_{0}^{1}
  \exp(4i{\RM_0}\disk\lambda^2s)
  \frac{\sin(2\Delta\RM\disk\lambda^2 s)}{2\Delta\RM\disk\lambda^2 s}\, ds \right|,
\end{equation}
where subscript zero refers to a value at the sector centre. We can
reasonably assume that depolarization due to Faraday dispersion in the
emitting layer is much stronger than Faraday dispersion in the
foreground screen (turbulent magnetic fields and thermal gas density
will be stronger near the mid-plane) and so $P_\mathrm{in}
P_\mathrm{ex}\sim P_\mathrm{in}$.

Since the galactic disk may be opaque to polarized emission at longer
wavelengths, mainly due to internal Faraday dispersion, the effective
path length can differ from that suggested by the disk scale height
(c.f.\ Sokoloff et al.\ \cite{Sokoloff98}, Berkhuijsen et al.\ 
\cite{Berkhuijsen97}).  We use the `opaque layer' approximation of
Sokoloff et al.\ (\cite{Sokoloff98}, Sect.~6.3) to describe the
visible depth $\Delta z$ in terms of the depolarization due to
internal Faraday dispersion assuming that all the observed polarized
emission at \wav{20} arises from an upper layer in the synchrotron
disk.  Figure~\ref{fig:diskfig} shows how $h\therm$, $h\syn$ and
$\Delta z$ are related. The path lengths over which the observed
polarized emission is produced are $\Delta z-(h\therm-h\syn)$ at
\wav{20} (here $\Delta z$ is a function of $\lambda$) and $2h\syn$ at
\wav{6}, where the disk is assumed to be transparent to polarized
emission.  Then a crude estimate of $\Delta z$ in terms of the
observed degrees of polarization follows from assuming that
depolarization due to internal Faraday dispersion is constant for all
sectors in a ring:
\begin{equation}
  \Delta z=h\therm+h\syn\left(2\,DP_\mathrm{in}-1\right)\;,
  \label{eq:Deltaz}
\end{equation}
where $DP_\mathrm{in}=P_\mathrm{in}(20\cm)/P_\mathrm{in}(6\cm)$ is
depolarization due to internal Faraday dispersion alone (see
Sect.~3.3.3 in Berkhuijsen et al.\ \cite{Berkhuijsen97}).  Since some
depolarization due to other mechanisms occurs within $\Delta z$, this
yields minimum values for the thickness of the visible layer.
Equation~(\ref{eq:Deltaz}) then gives the parameter $\xi_{\lambda}$ in
Eq.~(\ref{eq:xi}), in terms of $q=h\syn/h\therm$ and $DP_\mathrm{in}$,
via (Berkhuijsen et al.\ \cite{Berkhuijsen97})
\begin{equation}
  \xi_{\lambda}=\frac{1}{2}\left(1+\frac{\Delta z -h\syn}{h\therm}\right),
  \label{eq:xideltaz}
\end{equation}
so we have
\begin{equation}
  \xi_{\lambda}=1+q(DP_\mathrm{in}-1).
  \label{eq:xiq}
\end{equation}
The parameter $\xi_{\lambda}$ links the modelled regular magnetic
field described in Sect.~\ref{sec:model}, from which we obtain $\RM$
and $\Delta \RM$, and the model for Faraday depolarization described
in this section. Our aim is to obtain satisfactory fits for both
models \emph{using the same $\xi_{\lambda}$ in each}.

For a thermal layer thicker than the synchrotron disk (the
configuration that produces a foreground Faraday screen), $\Delta z
\geq h\therm-h\syn$, Berkhuijsen et al.\ (\cite{Berkhuijsen97}) showed
that
\begin{eqnarray}
  {\RM_0}\disk & = & {\textstyle\frac{1}{2}}\RM_0
        \frac{\Delta z - (h\therm-h\syn)}{h\therm}=q\RM_0\, DP_\mathrm{in},
  \label{eq:RMdisk}\\
  {\RM_0}\screen & = & \RM_0
        \frac{h\therm-h\syn}{h\therm}=(1-q)\,\RM_0,
  \label{eq:RMscreen}
\end{eqnarray}
where the final equalities result from substitution of
Eq.~(\ref{eq:Deltaz}). Using Eqs.~(\ref{eq:RMdisk}) and
(\ref{eq:RMscreen}) we can express $P(\lambda)$ in
Eq.~(\ref{eq:DPtot}) as a function of $q=h\syn/h\therm$, the ratio of
the scale heights of the synchrotron and thermal disks, and
$DP_\mathrm{in}$.  We fit the values of $q$ and $DP_\mathrm{in}$ by
comparing $\DP_{20/6}$ obtained from Eq.~(\ref{eq:DPtot}) to the
observed values in each sector. Figure \ref{fig:DPRing3} compares the
calculated $\DP_{20/6}$, for $q$ between $0.1$ and $0.4$, with
observed values in the ring 10--12 kpc, and shows how increasing the
relative scale height of the synchrotron disk reduces the effect of
the Faraday screen and enhances depolarization due to differential
Faraday rotation, which is strongest on the major axis.

\begin{figure}
  \includegraphics[width=0.45\textwidth]{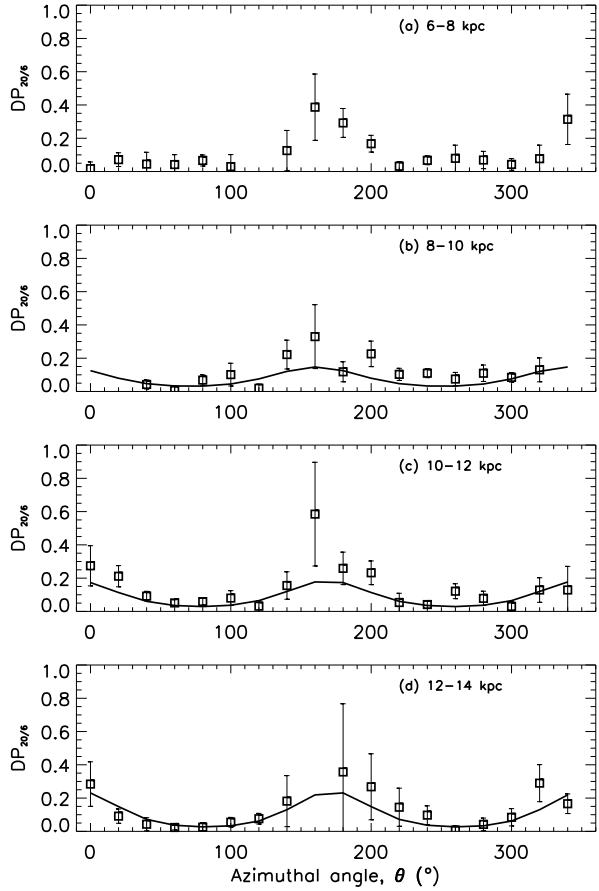}
  \caption{Observed (squares with error bars representing $1\sigma$ error)
    and expected (solid line) depolarization between \wav{20} and
    \wav{6}, $\DP_{20/6}$, due to Faraday effects using
    Eq.~(\ref{eq:DPtot}), (\ref{eq:RMdisk}) and (\ref{eq:RMscreen}),
    for each of four radial rings.  The expected depolarization is
    fitted by varying the ratio of the synchrotron to thermal disk
    scale heights, $q=h\syn/h\therm$ and depolarization due to
    internal Faraday dispersion, $DP_\mathrm{in}$.  The fitted values
    of $q$ are $0.6$, $0.6$, $0.4$, and $0.3$ and of $DP_\mathrm{in}$
    $0.1$, $0.1$, $0.3$ and $0.2$ in panels (a), (b), (c) and (d),
    respectively. The solid line was obtained using RM derived from
    the regular magnetic field fits of Table~\ref{tab:results}.}
  \label{fig:DPAllRing}
\end{figure}

\begin{table}
\caption{Parameters derived from the model of Faraday depolarization
and their $2\sigma$ errors.}
\begin{center}
\begin{tabular}{llll} \hline \noalign{\smallskip}
  Ring (kpc) & $q=h\syn/h\therm$ & $DP_\mathrm{in}$ & $h\therm$ (kpc)  \\
  \hline 
  6--8 & \multicolumn{3}{c}{No statistically good fit found}  \\[3pt]
  8--10 & $0.6\pm 0.1$ & $0.1\pm 0.1$ & $0.55\pm0.10$ \\[3pt]
  10--12 & $0.4^{+0.4}_{-0.2}$ & $0.3^{+0.7}_{-0.1}$ & $0.90^{+0.90}_{-0.45}$ \\[3pt]
  12--14 & $0.3^{+0.5}_{-0.3}$ & $0.2\pm 0.2$ & $1.3^{+2.6}_{-0.8}$ \\[3pt]
  \hline \\
  \label{tab:dp}
\end{tabular}
\end{center}
\end{table}


Figure \ref{fig:DPAllRing} shows the azimuthal variation of the
depolarization for the best fitting $q$ for each ring. The fitted
values of $q=h\syn/h\therm$ and $DP_\mathrm{in}$ are given in
Table~\ref{tab:dp}, where the quoted $2\sigma$ errors represent the
extent of the $(q, DP_\mathrm{in})$-parameter space within the
relevant $\chi^2$ contour. These errors are large, but it is
remarkable that we can successfully model Faraday depolarization in
such a complex system using such a simple, two parameter model.
Despite the uncertainty in the precise values of the model parameters,
the key result of this section is robust; the strong azimuthal pattern
of depolarization can only be explained by a Faraday screen acting
within M\,31 and hence the thermal electron layer must be
significantly thicker than the synchrotron emitting layer.

For $8<r<14 \kpc$ the modelled $\DP_{20/6}$ reproduce the observations
well. For each of these rings the fitted $\DP_{20/6}$ meets the
$\chi^2$ test for statistical significance at the $2\sigma$ level. In
the two rings at largest radii the results of the depolarization
modelling are \emph{fully consistent} with the polarization angle
model used to deduce the regular magnetic field structure in
Sect.~\ref{sec:model}. The two models are linked by the parameter
$\xi_{\lambda}$ -- a weighting for the depth in the emission layer
visible at long wavelengths -- in Eqs.~(\ref{eq:xi}) and
(\ref{eq:xiq}). For the rings 10--12~kpc and 12--14~kpc, $q$ and
$DP_\mathrm{in}$ in Table~\ref{tab:dp} give $\xi_{20}=0.7\pm{0.3}$ and
$\xi_{20}=0.8^{+0.2}$, respectively. In Sect.~\ref{sec:model} we
adopted $\xi_{20}=0.75$ for all of the rings and discrepancies of
order $\pm 0.1$ have a negligible effect on the fitted magnetic field
parameters given in Table~\ref{tab:results}.

For the ring 8--10~kpc, $q$ and $DP_\mathrm{in}$ give $\xi_{20}=0.5\pm
0.2$, whereas the best fit to the polarization angles in
Table~\ref{tab:results} requires $\xi_{20}=0.75$. We can achieve self
consistency between the magnetic field and depolarization models by
discarding more measured polarization angles in Sect.~\ref{sec:model},
i.e.\ by making the model of the magnetic field worse. However, the
main problem with the depolarization model in this ring is that around
the north end of the major axis our method of averaging the data gives
zero average polarized emission at \wav{20} in three sectors. (The
polarized intensity is averaged from maps smoothed to $3\arcmin$
resolution; the polarization angles are derived from $45\arcsec$ maps
at \wav{20} and do not suffer from this problem.) Without measurements
at \emph{both} ends of the major axis the fitted $\DP_{20/6}$ favours
a model with stronger Faraday dispersion i.e.\ a model that has a less
prominent double minimum. For these reasons we prefer to retain the
magnetic field model obtained with $\xi=0.75$ -- to keep $\xi_{20}$
the same for each ring -- and accept that the depolarization model for
this ring is poorer than for 10--12~kpc and 12--14~kpc.

The quality of the fit is clearly bad in the ring 6--8\,kpc
(Fig.~\ref{fig:DPAllRing}a) where the magnetic field, deduced in
Sect.~\ref{sec:model}, contains both the axisymmetric ($m=0$) and the
quadrisymmetric ($m=2$) components as given in
Table~\ref{tab:results}. In Sect.~\ref{subsec:model:ring1} we show
that this ring may have a more complicated regular magnetic field
structure than the purely azimuthal fields in the other rings.  The
modelled azimuthal patterns of RM and the gradient in RM are rather
complicated in the ring 6--8\,kpc and no good fit can be obtained. The
results would be better if we used a simpler fit involving a purely
axisymmetric magnetic field. However, as explained in
Sect.~\ref{subsec:model:ring1}, nine polarization angle measurements
must be discarded to make an $m=0$ magnetic field model, and the
consequent degrading of the regular magnetic field model is not
justified.

Using $h\syn$, at \wav{20}, from Table~\ref{tab:syndisk}, we can
estimate $h\therm$ from the fitted values for $q=h\syn/h\therm$. The
results are shown in Table~\ref{tab:dp}. These scale heights are about
a factor of two or more greater than previously expected in M\,31,
where the low star formation rate and absence of a radio halo were
thought to imply the likely absence of a thick ionized disk (Walterbos
\& Braun \cite{Walterbos94}).

We emphasize that the gradients in rotation measure producing most of
the depolarization in M\,31 are due to the highly axisymmetric regular
magnetic field that we find from an analysis of polarization angles in
Sect.~\ref{sec:model}. For simplicity, in modelling the depolarization
we assumed that the regular magnetic field has the same configuration
and strength throughout the full vertical extent of the thermal layer
(including the synchrotron emitting disk). If the regular magnetic
field strength or the thermal electron density has a maximum above the
emitting disk (i.e., at $z\gtrsim 300\p$), the RM required to produce
the observed depolarization can be generated in a thinner layer and
$h\therm$ will be lower than estimated above, but still
$h\therm>h\syn$.

In Sect.~\ref{subsubsec:pi} the limitations of the \wav{20}
polarization data when smoothed to $3\arcmin$ were discussed;
foreground emission from the Milky Way cannot be subtracted from the
emission from M\,31 and so the \wav{20} polarized intensities are
upper limits. The values of $q=h\syn/h\therm$ and $DP_\mathrm{in}$
shown in Table~\ref{tab:dp} were derived assuming that \emph{all} of
the polarized emission at \wav{20} comes from M\,31 and so are
\emph{upper limits} on $q$ and $DP_\mathrm{in}$.

The corresponding \emph{lower limits} (i.e., giving depolarization
stronger than required) can be obtained assuming that the emission
from M\,31 is nearly completely depolarized at \wav{20}.  Without
Faraday depolarization, the \wav{20} polarized emission from M\,31
will have the same azimuthal pattern as the \wav{6} $PI$ shown in
Fig.~\ref{fig:RM}(c). Total depolarization ($\DP_{20/6}=0$) will occur
when $DP_\mathrm{in}$ and differential rotation are just strong enough
to depolarize the emission on the major axis ($\theta=0\degr,
180\degr$) and RM gradients are just sufficient to depolarize emission
from the minor axis ($\theta=90\degr, 270\degr$). From
Fig.~\ref{fig:DPRing3}(a) we estimate that, for the ring
10$<r<$12~kpc, complete depolarization will occur if $q\simeq 0.1$ and
$DP_\mathrm{in}\simeq 0.1$. These are the lower limits on $q$ and
$DP_\mathrm{in}$.

The regular magnetic field must be coherent in $z$ over at least the
scale height of the thermal disk, and we have shown that the latter
must exceed that of the synchrotron disc.  This poses the intriguing
question of why \emph{the cosmic rays in M\,31 are confined to a layer
several times thinner than the regular magnetic field}. One possible
answer relies on the usual assumption of equipartition between the
cosmic ray and magnetic field energy densities. Then the synchrotron
emissivity depends upon the fourth power of the magnetic field and so
$h_{B}\sim 4 h_{syn}\simeq 1.5\kpc$. This scale height is in good
agreement with $h\therm$ derived from our analysis of depolarization
(at least for the two rings with the most reliable model of
$\DP_{20/6}$). In M\,31, the magnetic field is well ordered with
$\Br\simeq\bt$ and there is no significant vertical component of the
magnetic field (see Sect.~\ref{sec:model}). This may be sufficient to
suppress diffusion of cosmic rays perpendicular to the disk plane and
so constrain them to the same layer as their sources.

\subsection{Thermal electron densities}
\label{subsec:ne}
Using Eq.~(\ref{eq:FRM}) and the equipartition regular magnetic field
strengths given in Table~\ref{tab:syndisk}, the rotation measures from
Table~\ref{tab:results} and the thermal disk scale heights of
Table~\ref{tab:dp} we can derive average thermal electron densities
for M\,31 in the radial range $8<r<14\kpc$. This gives $\langle
n_\mathrm{e}\rangle\simeq 0.008,0.007$ and $0.004\cmcube$ for the
rings 8--10, 10--12 and 12--14~kpc, respectively.  These values refer
to the upper layers of the thermal electron layer, $z\ga
h_\mathrm{syn}\simeq200$--300\,pc, that act as the Faraday screen.

Electron density closer to the midplane can be obtained from the
amount of depolarization due to Faraday dispersion between \wav{20}
and \wav{6}, $DP_\mathrm{in}\simeq0.1$ as obtained above. Using
Eq.~(\ref{eq:DPin}) with $b=5\mkG$, $L=200\p$ and $d=50\p$, we obtain
$\sigma_\mathrm{RM}\simeq550\left(\langle
  n_\mathrm{e}^2\rangle^{1/2}/1\cmcube\right)\FRM$ and then
$DP_\mathrm{in}=0.1$ corresponds to $n_\mathrm{e}\simeq0.1\cmcube$.
This estimate is compatible with that obtained by Walterbos \& Braun
(\cite{Walterbos94}) from \ion{H}{$\alpha$} emission measures of the
diffuse ionised gas, $\langle n_\mathrm{e}\rangle\simeq
0.08$--$0.04\cmcube$ with a filling factor 0.2.

Thus, the equipartition magnetic field strength, rotation measures and
the scale heights of the thermal disk derived in our models produce an
estimate for $\langle n_\mathrm{e}\rangle$ that is in broad agreement
with $\langle n_\mathrm{e}\rangle$ obtained from completely different
data and methods.

Berkhuijsen et al.\ (\cite{Berkhuijsen03}) note that there is little
correlation between RM and thermal emission in M\,31 and suggest that
the small filling factor of \ion{H}{ii} regions may be the reason.
This is consistent with our conclusion that much of the Faraday
rotation in M\,31 is produced in a Faraday screen.

\subsection{Review of the method}
\label{subsec:methdiscuss}
In fitting the modelled to observed polarization angles in
Sect.~\ref{sec:model}, we use the parameter $\xi_{\lambda}$ to account
for the partial opacity of the galaxy's disk to polarized emission at
\wav{20}. In order to estimate $\xi_{\lambda}$ we need to know the
ratio of the scale heights of the synchrotron and thermal disks,
$q=h\syn/h\therm$, but the values for $q$ deduced in
Sect.~\ref{subsec:depolar:model} make use of RM calculated from the
fits of Sect.~\ref{sec:model}. We used an iterative approach to try to
obtain a model consistent with both the observed depolarization and
polarization angles. This method was successful for the two outer
rings, after one iteration, but not for the rings 6--8~kpc and
8--10~kpc. For these rings we adopted $\xi=0.75$ from the
self-consistent models of the rings 10--12~kpc and 12--14~kpc.

\section{Summary}
\label{sec:summary}

Sensitive, high resolution, multi-wavelength radio polarization
observations have been used to study the magnetic field of M\,31,
between the radii of $6$ and $14\kpc$. The powerful method of using
polarization angles to uncover the regular magnetic field structure
was supplemented by a systematic analysis of depolarization to produce
a model of the regular magnetic field which is consistent with
\emph{all} of the radio polarization data for $10<r<14\kpc$.

Our main conclusions are as follows :

1. The regular magnetic field in M\,31 is axisymmetric to a very good
approximation.

2. The magnetic field has a significant radial component at all radii
and so is definitely not purely azimuthal. Vector lines of the regular
magnetic field in the radial range $6\lesssim r \lesssim 10\kpc$ can
be approximated by trailing logarithmic spirals, with the pitch angle
$p\simeq -16\degr$. The magnetic spiral becomes tighter at large
radii, with $|p|$ decreasing to $p\simeq-7\degr$ at $r=12$--$14\kpc$.
The magnitude and trend of magnetic pitch angles is in broad agreement
with those expected from dynamo theory (Shukurov\ \cite{Shukurov00}).

3. Analysis of the azimuthal pattern of the wavelength dependent
depolarization reveals that a Faraday active screen lies above the
synchrotron emitting disk of M\,31. The diffuse thermal disk is
thicker than previously expected, with a scale height of $h\therm\sim
1\kpc$.

4. The scale height of the regular magnetic field is at least equal to
$h\therm$.

5. The magnetic field in M\,31 extends inside and outside of $r\simeq
10\kpc$, as found by Han et al.~(\cite{Han98}), and does not have a
strong maximum at this radius. The bright radio ring is a result of a
high density of cosmic ray electrons.

6. The equipartition field strengths are about 5$\mkG$ for both the
regular and turbulent field components, without significant variation
between 6~kpc and 14~kpc radius.

7. Faraday rotation measures and equipartition field strengths are in
agreement for average electron densities of 0.008--0.004$\cmcube$.
The electron densities inferred from the Faraday dispersion measures
are $\simeq 0.1\cmcube$, close to the average electron densities found
by Walterbos \& Braun (\cite{Walterbos94}). This suggests that the
diffuse ionised gas is mainly responsible for the Faraday rotation,
with little contribution to RM from \ion{H}{ii} regions.

Our analysis of depolarization in M\,31 is the most extensive
undertaken to date for a spiral galaxy, and shows that the theory of
radio depolarization developed by Burn (\cite{Burn66}) and Sokoloff et
al.\ (\cite{Sokoloff98}) can be used not only to identify the causes
of depolarization, but also to reveal properties of the diffuse ISM in
external galaxies.

\begin{acknowledgements}
  We thank Marita Krause for comments following careful reading of the
  manuscript.
  
  Our results are based on observations with the Effelsberg 100m
  telescope of the MPIfR.
  
  AF was funded by a PPARC studentship at the University of Newcastle,
  where much of this work was undertaken. Financial support from NATO
  (Grant PST.CLG 974737), PPARC (Grant PPA/G/S/2000/00528) and a
  University of Newcastle Small Grant are gratefully acknowledged.
\end{acknowledgements}

\appendix
\section{Depolarization mechanisms}
\label{app}

\begin{figure}[h]
  \includegraphics[width=0.45\textwidth]{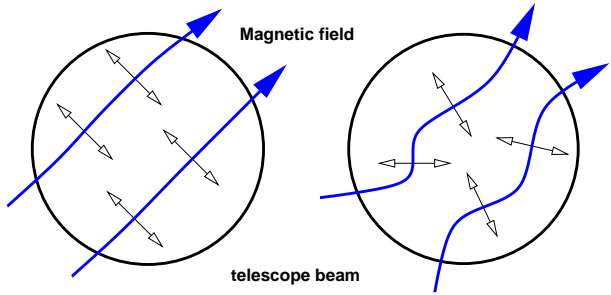}
  \caption{Sketch showing wavelength independent depolarization.
     The double headed arrows represent E-vectors.}
  \label{fig:inddp}
\end{figure}

\textbf{Wavelength independent depolarization} is caused by tangling
of magnetic field lines in the emitting region (Figure~\ref{fig:inddp}).
The intrinsic polarization angle of synchrotron radiation is
perpendicular to the local magnetic field orientation and so tangled
magnetic field lines result in emission at a range of polarization
angles within a single telescope beam. As long as the beam sizes are
equal, the degree of depolarization due to tangled magnetic field
lines will be the same at all wavelengths.

\begin{figure}[h]
  \includegraphics[width=0.45\textwidth]{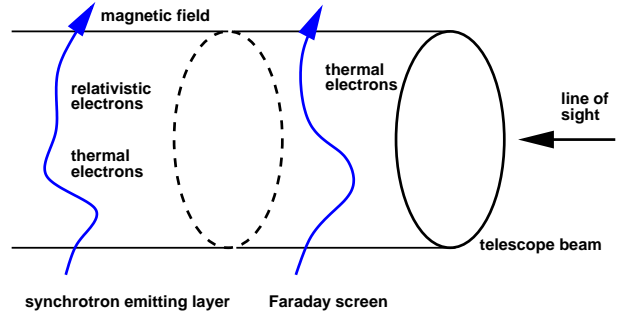}
  \caption{Sketch of synchrotron emitting layer and a foreground Faraday screen.}
  \label{fig:layersdp}
\end{figure}

Faraday rotation by both regular and turbulent magnetic fields results
in \textbf{wavelength dependent depolarization.} It is useful to
consider separately Faraday effects within the synchrotron emitting
layer and Faraday rotation in regions where there is no emission, i.e.\
within a Faraday screen (Figure~\ref{fig:layersdp}).

\begin{figure}[h]
  \includegraphics[width=0.45\textwidth]{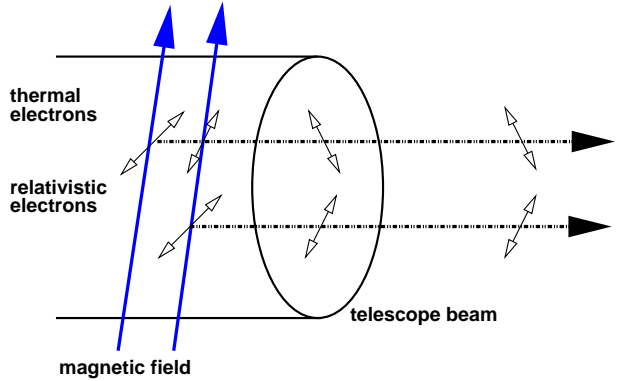}
  \caption{Differential Faraday rotation
    occurs within the synchrotron emitting layer. Emission from
    different depths \emph{along the same line of sight} undergoes
    different amounts of Faraday rotation, causing depolarization. For
    clarity the sketch separates emission from different depths.}
  \label{fig:diffrot}
\end{figure}

The regular field in the synchrotron emitting layer causes
depolarization by \textbf{differential Faraday rotation}, whereby
polarized emission from different depths along the line of sight is
rotated by different amounts (Figure~\ref{fig:diffrot}). In a slab
with uniform magnetic field and electron density the degree of
polarization is (Burn \cite{Burn66}, Sokoloff et al.
\cite{Sokoloff98})
\begin{equation}
  \label{eq:DPreg}
  P_\mathrm{reg}=P_0
  \left\vert \frac{\sin(2\RM\lambda^2)}{2\RM\lambda^2} \right\vert,
\end{equation}
where $\RM= 0.81 \langle n_\mathrm{e}\rangle \Brpa L/2$ is the
observed rotation measure in units of $\!\radm$, with $\langle
n_\mathrm{e}\rangle$ the average thermal electron density in
$\!\cmcube$, $\Brpa$ the component of $\Br$ parallel to the line of
sight in $\mkG$ and $L$ the path length through the Faraday active
emitting layer, in $\!\p$.

\begin{figure}[h]
  \includegraphics[width=0.45\textwidth]{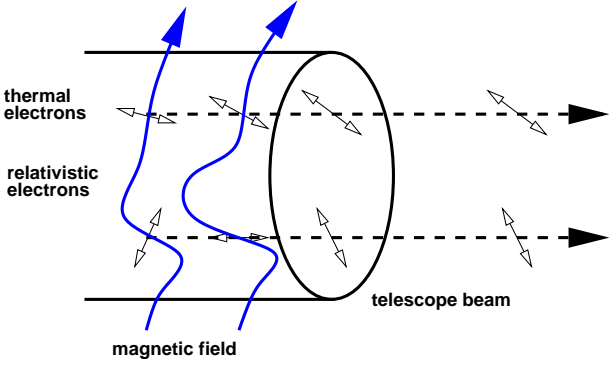}
  \caption{Sketch illustrating Faraday dispersion.
    When this effect occurs within the synchrotron emitting layer it
    is called internal Faraday dispersion; occurrence in a Faraday
    screen (no synchrotron emission) is external Faraday dispersion.
    Note that the required variation of rotation measure within the
    beam could also be caused by fluctuations of thermal electron
    density.  In that case the magnetic field could be totally regular
    but Faraday dispersion would still occur.  }
  \label{fig:disp}
\end{figure}

The presence of unresolved, turbulent magnetic field means that
polarized emission along different lines of sight within the telescope
beam undergoes different amounts of Faraday rotation
(Figure~\ref{fig:disp}).  When the emitting and rotating layers
coincide, the effect is called \textbf{internal Faraday dispersion},
and the degree of polarization is given by Sokoloff et al.\
(\cite{Sokoloff98}) as
\begin{equation}
  \label{eq:DPin}
  P_\mathrm{in}=P_0
  \frac{1-\exp(-2\sigma_\mathrm{RM}^2 \lambda^4)}{2\sigma_\mathrm{RM}^2 \lambda^4},
\end{equation}
where $\sigma_\mathrm{RM}^2=0.81\langle n_\mathrm{e} \bt\rangle^2 2L d$
with $d$ the correlation scale (half the turbulent cell size) of the
turbulent magnetic field in parsecs. Burn (\cite{Burn66}) gives the
depolarization due to \textbf{external Faraday dispersion} (due to
turbulent fields in front of the synchrotron source) as
\begin{equation}
  \label{eq:DPex}
  P_\mathrm{ex}=P_0\exp(-2\sigma_\mathrm{RM}^2 \lambda^4).
\end{equation}

\begin{figure}[h]
  \includegraphics[width=0.45\textwidth]{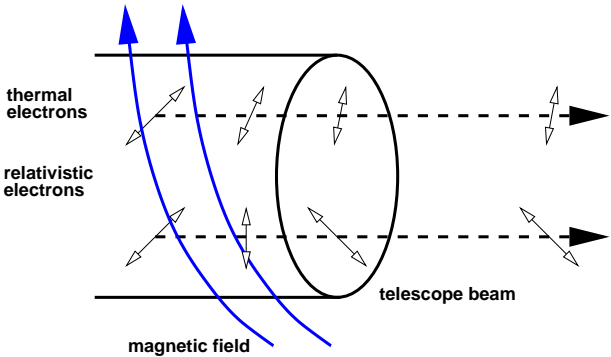}
  \caption{Depolarization due to gradients in rotation measure:
    if RM varies across the beam, different parts of the beam area
    contribute with different polarization angles.  Note that
    variations of thermal electron density or magnetic field strength
    can also produce RM gradients even in a totally regular magnetic
    field. }
  \label{fig:rotgrad}
\end{figure}

\textbf{Gradients in rotation measure} across the beam cause
depolarization that is especially strong when the resolution of
observations is low (Figure~\ref{fig:rotgrad}).  For depolarization by
RM gradients \textbf{within the synchrotron source}, including the
effect of differential Faraday rotation, Sokoloff et al.~(\cite{Sokoloff98})
obtained
\begin{equation}
  \label{eq:DPgradint}
  P_\mathrm{\Delta in}= P_0\left\vert \int_{0}^{1}
  \exp(4i\,\mathrm{RM_0}\, \lambda^2s)
  \frac{\sin(2\,\Delta \RM\, \lambda^2 s)}
  {2\, \Delta\RM\, \lambda^2 s} \,ds \right\vert,
\end{equation}
where $\mathrm{RM_0}$ is $\RM$ at the centre of the beam, $\Delta\RM$
is the increment in $\RM$ across the beam and the normalized
integration variable $s$ describes the line of sight within the
synchrotron disk, $0\leq s \leq 1$.  Equation~(\ref{eq:DPgradint})
holds for both resolved and unresolved RM gradients (Sokoloff et al.
\cite{Sokoloff98}). Depolarization by a \textbf{resolved RM gradient
  in a Faraday screen} is given by Sokoloff et al.~(\cite{Sokoloff98})
as
\begin{equation}
  \label{eq:DPgradext}
  P_\mathrm{\Delta ex} = P_0\left\vert \exp\left[2i\,\mathrm{RM_0}\, \lambda^2-
  2\left(\Delta\RM\, \lambda^2\right)^2\right] \right\vert.
\end{equation}


\end{document}